\documentclass[sigconf, anonymous=false]{acmart}

\usepackage{booktabs} 
\usepackage[caption=false,font=footnotesize]{subfig}
\usepackage{multirow}
\usepackage{mathtools, nccmath, textcomp}
\usepackage{balance}
\graphicspath{{fig/}}

\copyrightyear{2019}
\acmYear{2019}
\setcopyright{acmlicensed}
\acmConference[-]{-}
\acmBooktitle{-}
\acmPrice{-}
\acmDOI{-}
\acmISBN{-}

\begin{document}
\title[A Multivariate Appliance Event Detection]{Appliance Event Detection - A Multivariate, Supervised Classification Approach \vspace{2mm}}

\author{Matthias Kahl, Thomas Kriechbaumer, Daniel Jorde, Anwar Ul Haq, and Hans-Arno Jacobsen}
\affiliation{%
  \department{Chair for Application and Middleware Systems}
  \institution{Technische Universität München, Germany}
}
\email{matthias.kahl@in.tum.de}

%
%
%
%

\renewcommand{\shortauthors}{M. Kahl et al.}


\begin{abstract}
Non-intrusive load monitoring (NILM) is a modern and still expanding technique, helping to understand fundamental energy consumption patterns and appliance characteristics. Appliance event detection is an elementary step in the NILM pipeline. Unfortunately, several types of appliances (e.g., switching mode power supply (SMPS) or multi-state) are known to challenge state-of-the-art event detection systems due to their noisy consumption profiles. Classical rule-based event detection system become infeasible and complex for these appliances. By stepping away from distinct event definitions, we can learn from a consumer-configured event model to differentiate between relevant and irrelevant event transients.

We introduce a boosting oriented adaptive training, that uses false positives from the initial training area to reduce the number of false positives on the test area substantially. The results show a false positive decrease by more than a factor of eight on a dataset that has a strong focus on SMPS-driven appliances. To obtain a stable event detection system, we applied several experiments on different parameters to measure its performance. These experiments include the evaluation of six event features from the spectral and time domain, different types of feature space normalization to eliminate undesired feature weighting, the conventional and adaptive training, and two common classifiers with its optimal parameter settings. The evaluations are performed on two publicly available energy datasets with high sampling rates: BLUED and BLOND-50.

\end{abstract}

%
%



\maketitle


\section{Introduction}


Natural energy resources are depleting at an alarming rate and, at the same time, the demand for energy is steadily increasing. Recently, many approaches have been proposed to decrease our reliance on these resources by increasing energy efficiency. Non-intrusive load monitoring (NILM) provides detailed information on the energy consumption for consumers in residential or industrial areas. Surveys indicate that appliance-level consumption feedback can increase consumers awareness to energy wasting and therefore reduce their energy consumption \cite{Kelly2016}. NILM is an intelligent energy monitoring technique which utilizes a single energy monitor to retrieve information of appliances from aggregated loads, such as power consumption and appliance type, in a non-intrusive way. Apart from energy savings, NILM is a helpful tool for predictive maintenance \cite{Abdelgayed2018} or the determination of motor speed \cite{Orji2015}.

\begin{figure}[htbp]
\centering
\subfloat{\includegraphics[width=0.95\linewidth]{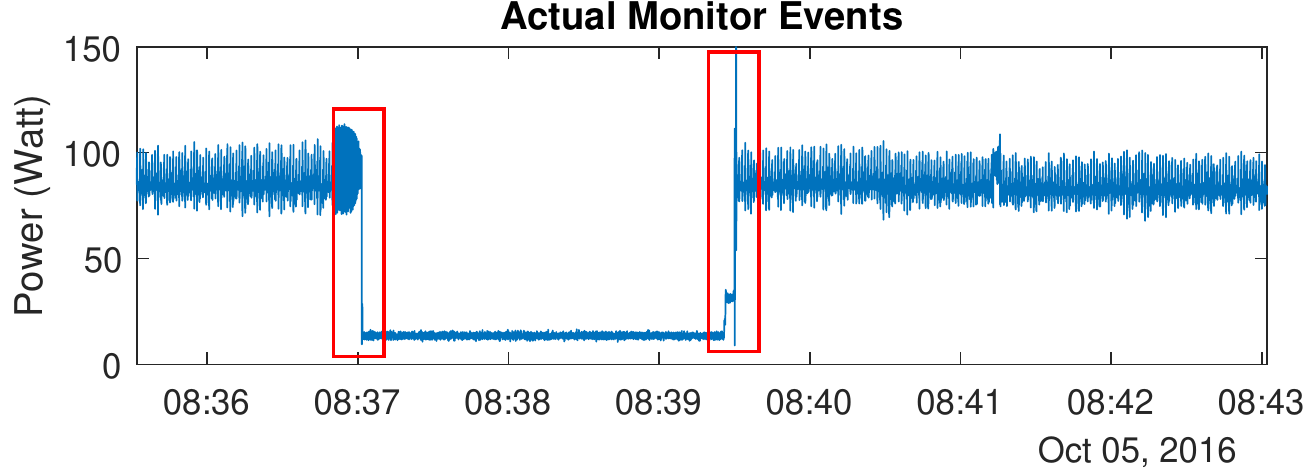}} \\
\subfloat{\includegraphics[width=0.95\linewidth]{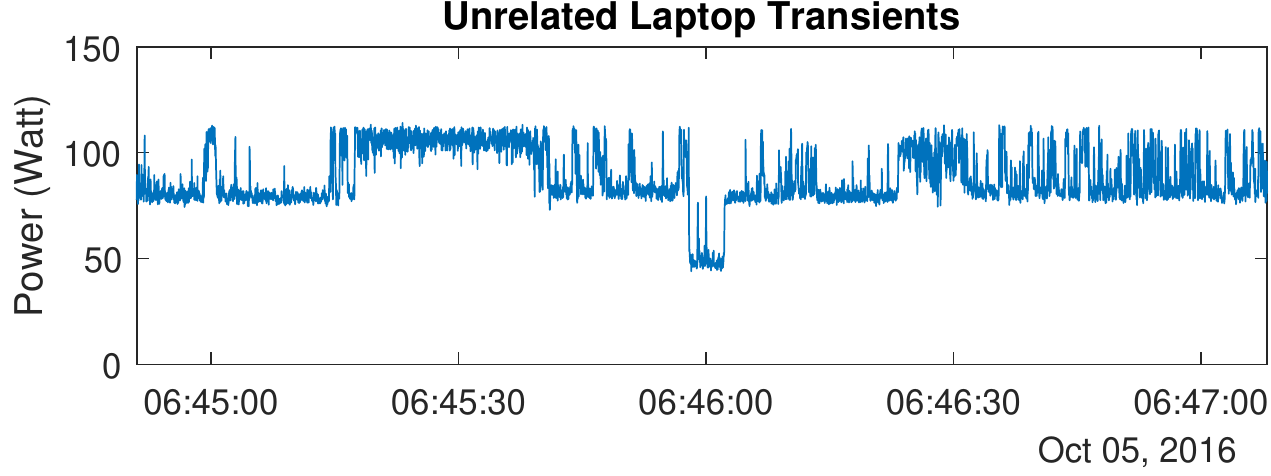}}
\caption{The first plot shows an actual OFF, followed by an ON event of a monitor. The second plot shows sudden laptop transients, that most likely stem from processor load changes. The goal is to differentiate between actual ON / OFF events and transients that are irrelevant to the user.}
\label{fig:eventsVStransients}
\Description{Two plots of a monitor and a laptop event.}
\end{figure}

Appliance events play an essential role in NILM since these events are the time points where the energy consumption significantly changes. Therefore, it is an important task to identify appliance events for all following steps correctly. Regarding NILM, an appliance event is often defined as the transition between two steady states of a time series \cite{Wild2015}. These time series are mostly in the form of current, real or reactive power, and voltage distortion \cite{Meziane2017}. Other metrics, such as admittance or derivation of current, allow identifying appliance events as well, but seem to be not considered for now. Several event-based approaches have been proposed to retrieve detailed consumption information from aggregated load signals \cite{Hart1992, Armel2013, Batra2014}. Event-based NILM approaches differ in performance, based on the number and types of appliances, the sampling frequency of the acquired data, the quality of the event metrics, and the complexity of utilized disaggregation algorithms. These factors determine the detection accuracy of the appliances from the aggregated load.

A considerable amount of inaccuracies in NILM disaggregation stems from the event detection, which depends mostly on the observed appliance type. A high false positive rate is a common problem in event detection. Events from resistive appliances are usually steep, undistorted, and easily detectable due to very low noise in their steady-state consumption. SMPS-driven continuous load appliances (desktop PCs, laptops, LED-TVs, etc.) on the other hand can draw strong event-like transients mid-usage that satisfy typical event rules but do not match any physical state change or user interaction (see Figure~\ref{fig:eventsVStransients}).

Event detection algorithms are often using a rule-based system with hand-crafted and empirically selected sets of rules \cite{Girmay2016, Trung2014, Jin2011}. From a certain point of rule complexity and due to the presence of manually labeled data, a supervised learning approach is worth to consider \cite{IanGoodfellow2016}. The high noise and variances in the current waveform of SMPS-driven appliances is hardly processable with rule-sets.

Therefore, for our approach, we replace hand-crafted rules with a multivariate, binary classification to distinguish between unrelated event-like transients and actual user relevant appliance events. Our classification system learns from customer-labeled events to distinguish between appliance ON / OFF events and customer irrelevant event-like transients in \emph{events} and \emph{non-events}. As a result of our critical discussion about hard-coded appliance event definitions, our event detection is designed to retrieve a flexible event definition supervised from representative examples. Our two-step adaptive learning approach that is oriented on the boosting algorithm \cite{Geron2017}, ensures a relevant selection of training samples for the event and non-event class by learning from false positives. Our experiments show that the algorithm can reduce the number of unrelated event-like transients (false positives) significantly.

The rest of the paper is organized as follows: Section 2 gives an overview of related work and discusses several event definitions. A detailed description of the multivariate event detection approach can be found in Section 3, while our experiments are given in Section 4. The results in Section 5 are consolidated and put into the NILM context in our conclusions in Section 6.



\section{Background and Related Work}

\begin{figure}[htbp]
\centering
\includegraphics[width=0.95\linewidth]{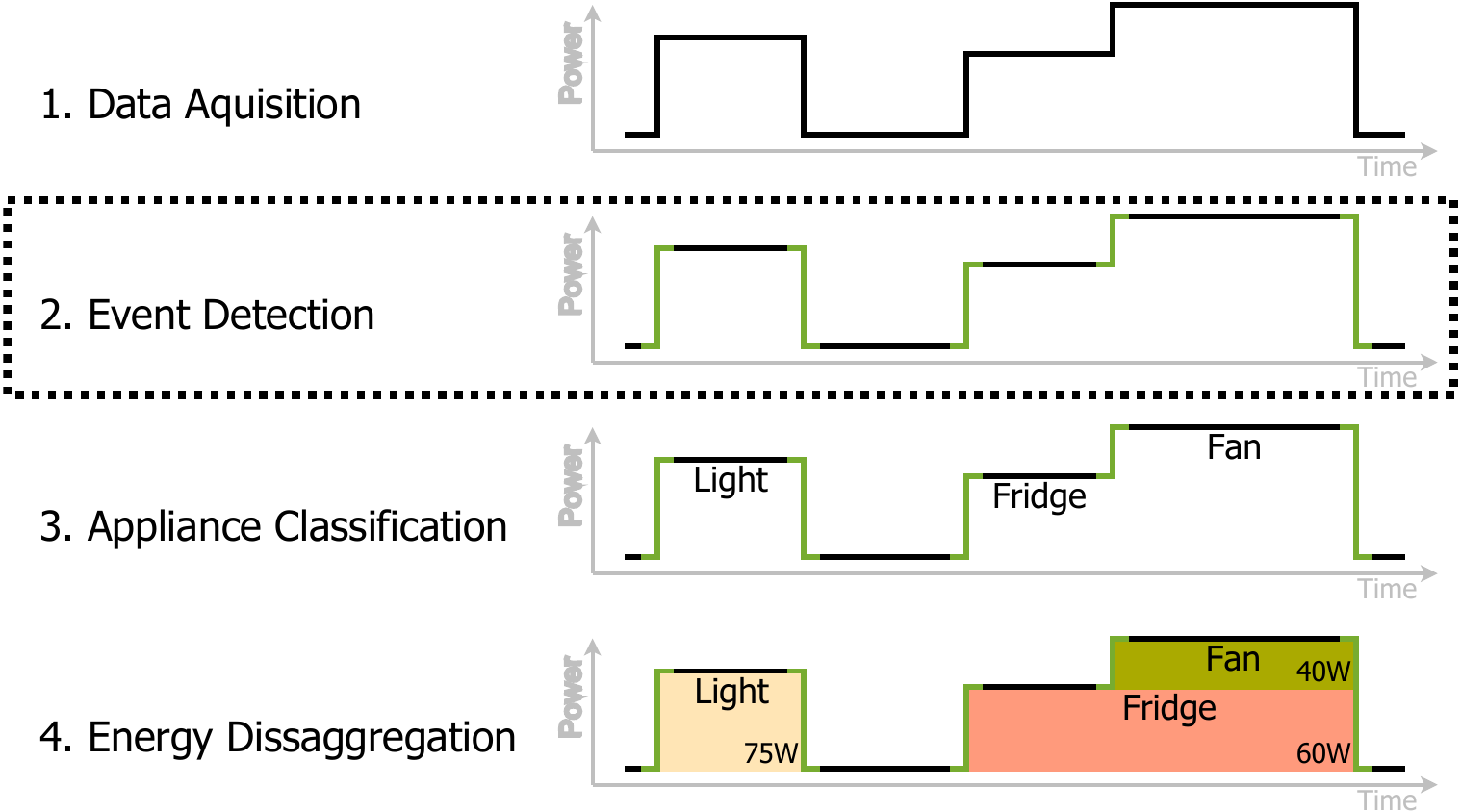}
\caption{The focus in this work is on event detection of the general NILM pipeline.}
\label{fig:EventExtraction}
\Description{The 4 steps of the NILM pipeline.}
\end{figure}

According to \citet{Anderson2012a} the elementary steps for energy consumption feedback with NILM are: (1) Signal measurement, (2) Appliance Event detection, (3) Appliance Event classification, and (4) Energy Disaggregation (see Figure~\ref{fig:EventExtraction}).

Multi-state and SMPS-driven appliances often show unrelated event-like transients due to appliance state changes. These transients can be caused, amongst others, by computers that switch spontaneously from idle to full processor load (see Figure~\ref{fig:eventsVStransients}). Organic-LED-driven monitors have an image dependent energy consumption that can switch from minimum load to maximum load in between milliseconds just by changing from black to white in the displaying image. These undesired or unrelated transients affect the appliance classification and power disaggregation performance and make the event detection a challenging part. Rule-based event detection algorithms would need a complex rule set that is hardly feasible and sensitive to environment changes or appliance set changes due to their inflexibility.

\subsubsection*{Event Detection}
NILM is commonly divided into event-based and state-based approaches. Event-based approaches rely on using detection algorithms in order to find electrical events such as switch-ON or switch-OFF of an individual appliance. State-based methods on the other hand, take into account every sample of the signal to perform the inference step. Event-based methods are generally more efficient in the inference step than state-based approaches. This efficiency is caused by pre-processing of the voltage and current signals with labeling and extracting the regions of interest of the signal after the events have occurred. \cite{Liang2010}. Most of the event-based methods rely on the switch continuity principle \cite{Makonin2016}, which was initially introduced by Hart in 1992 \cite{Hart1992}. It essentially states that there is only up to one event, i.e., not multiple ones, at a given point in time. Furthermore, it assumes that events are relatively rare when looking at the overall signal, allowing to see the event detection as anomaly detection. Sampling data at higher rates increases the validity of this principle. Employing this principle allows event-detection methods and other algorithms to treat electric events as being isolated from one another \cite{Makonin2016}.

Three categories of event detection approaches are introduced by \citet{Anderson2012a}. Expert heuristics describe mostly rule-based approaches that consider prior knowledge to define sets of parameters and thresholds \cite{Hart1992, Baranski2004}. Probabilistic models consider statistical metrics, including variance and standard deviation, to estimate the probability of a change in a time series \cite{Berges2011, Jin2013}. Approaches of the matched-filter category try to find a universal event pattern in the signal by exceeding a likelihood threshold \cite{Leeb1995, Shaw2008}. The approach of \citet{Anderson2012a} considers the usage of a modified general likelihood ratio detector to compare four different evaluation metrics.

\citet{DeBaets2016} apply a cepstrum smoothing high-pass filter to the signal. This way, only very low frequency and step changes remain in the signal. The assumption is that in the case of an event, all remaining low frequencies lie above a certain threshold. The optimal parameter values were empirically evaluated. De Baets et al. compare the results on the BLUED dataset with the chi-squared goodness-of-fit ($X^2$ GOF) approach by \citet{Jin2011a} and could reach comparable results.

\citet{Barsim2014} introduced an unsupervised event detection algorithm which creates the logarithm of the P,\,Q plane \cite{Hart1992} to find steady states as clusters, while transients are represented as single scatters or outliers. The extraction of actual events was performed in three stages: a coarse search, followed by a fine search, and a final verification stage. The unsupervised way has the advantage that no learning from existing ground truth is necessary. The results show a very similar performance compared to \citet{DeBaets2016}.

\citet{Wild2015} introduce a new event definition which gives events a dimension in time, they are not infinite anymore. This definition allows a Fisher discriminant analysis in combination with some constraints a robust unsupervised appliance event detection in the spectral domain.

\citet{Houidi2018} investigate three commonly used techniques for the abrupt event detection that are typically used in other research fields: the Effective Residual algorithm \cite{Berriri2012}, the Cumulative Sum (CUSUM) algorithm \cite{Trung2014}, and the Bayesian Information Criterion algorithm \cite{Ajmera2004}. These algorithms are probabilistic event detection techniques. By comparing the algorithms in a real-world environment, Houidi et al. conclude that the CUSUM algorithm outperforms the other two and achieves good results on their internal dataset.

\citet{Azzini2014} introduce the "window with margin" method. This threshold-based algorithm uses a sliding window and a subset of the samples within the window, i.e., samples from the beginning and the end of the window, to calculate two averages of the active power consumption. Azzini et al. then use heuristically defined thresholds to check if the difference between the averages exceeds a certain limit in order to detect events in the signal.

The event detection methods above are developed for residential settings, whereas \citet{Leeb1993} propose a multi-scale transient event detector for industrial settings. To tolerate overlapping events, the author's algorithm searches for time patterns of segments in the signal that exhibit significant variation instead of searching for complete transient shapes. The algorithm detects such segments by using a change-of-mean detector. The transient changes in the signal are then detected by using sets of the previously computed segments as features for particular events and a pattern matching algorithm.

In contrast to the majority of the event detection approaches, \citet{Cox2006} do not use current signals and analyze only aggregated voltage measurements. By using a spectral decomposition of the voltage signal to compute the harmonic voltage distortion, they are able to detect residential appliance events reliably. They further show that the voltage signal exhibits sufficient information to identify events.

All mentioned approaches have in common that all significant transients are interpreted as events. Every approach considers another event definition making it hard to compare their results. They do not allow to distinguish between different kinds of events or ignore undesired events.

\begin{table}[htbp]
\centering
\caption{Event detection results on BLUED, using different event definitions making the results hardly comparable}
\label{tab:relatedWork}
\begin{tabular}{lcc}
Work of\dots        & ~~ & \multicolumn{1}{c}{F-Score} \\ \hline \hline
\citet{DeBaets2016} & ~~ & 80.04                     \\
\citet{Jin2013}     & ~~ & 81.01                     \\
\citet{Wild2015}    & ~~ & 89.15                     \\ \hline
\end{tabular}
\end{table}

\subsubsection*{Datasets}
We use two common energy datasets to evaluate the introduced event detection algorithm. The Building-Level fUlly-labeled dataset for Electricity Disaggregation (BLUED) is being introduced by \citet{Anderson2012}. The dataset contains continuous voltage and current measurements of around one week from a single-family household. The aggregated consumption signal is measured in a high amplitude (16-bit) and temporal resolution (12\,kHz). To enable event detection research, Anderson et al. decided to label significant appliance state transients with timestamps and appliance information. The transient event ground truth stems from additional sensing such as light sensors and visual observation of humans. The resulting 1\,577 events of phase B are used for our experiments. An overview of event-detection results using the BLUED dataset can be seen in Table~\ref{tab:relatedWork}.

The Building-Level Office eNvironment dataset (BLOND) \cite{Kriechbaumer2018} contains long-term continuous measurements of a 3-phase energy supply to an office building. In this work, we use the appliance-level BLOND-50 sub-dataset, which contains mains voltage and current measurements of 90 observed sockets. The per-appliance electrical signals that aim for ground truth retrieval were collected with 6.4\,kHz and 12-bit resolution. The data amounts to 213 days of recording, from which we selected the period of November 2016 for all experiments on BLOND-50. The ground truth assigns an appliance type and the associated nameplate information with a monitored power socket.

\subsubsection*{Event Definition}
Regarding the event definition itself, multiple different interpretations of events can be found in the literature. \citet{Wild2015} present a classical and an extended event definition. A classical event is a "transient from one steady state to another steady state which definitely differs from the previous one" \cite{Wild2015}, while an extended event describes a "so-called active section where the signal is somehow deviating from the previous steady state" \cite{Wild2015}, which provides a higher resilience against peaks and short pulses. \citet{Anderson2012a} define an event with a state change of 30\,W for a certain amount of time in a concrete value-based way, while \citet{Jin2011} see event detection as a way to find ON and OFF transients of appliances. \citet{Girmay2016} see an event as an active region from any appliance activation in which the power consumption is "well above" the background power.

The list of definitions above shows that there is no common agreement on what an appliance event can be. The event detection performance depends strongly on the event definition itself. A simple definition that includes a significant change of power for a certain amount of time, regardless of the cause, can simply be put into a rule-based system that may allow for a perfect detection performance. From the consumer perspective, appliance ON / OFF events that have a causal origin (i.e., from user interaction or physical appliance state changes) are more relevant than transients that simply satisfy the rule set. In practice, the consumer might be interested in the fridge or washing machine spin cycles. The temporarily increased energy consumption from a laptop during an irregular 5 minute lasting operating system update or the suddenly content dependent energy consumption of an organic-LED-driven TV is only of minor interest to the consumer.

Our approach avoids a distinct, hard-coded appliance event definition by learning from individual consumer-configured appliance event segments to build a tailored event model. This way we step back from a distinct event definition in favor of a user-definable event model. Since events from different appliances show individual characteristics, a rule-based approach with thresholds may not be sufficient to find ON / OFF switches. Our system is able to learn from different event features in the time and spectral domain which are fed as features into a supervised binary classification system. To improve the classification performance, we introduce an adaptive training technique that learns from previously wrong detected transients that lie on the border between events and non-events.



\section{Multivariate Event Detection}


A reasonable appliance classification and disaggregation performance will only be achieved when the NILM system adapts to the deployed environment. The customization may include parameter settings of \emph{base load}, \emph{min/max appliance load} or \emph{max concurrent running appliances}. Besides those parameters, a consumer supervised appliance labeling for system training purposes, over a certain amount of time (e.g., few days/weeks), will result in considerably improved classification and disaggregation performance \cite{Kahl_2017b}. 


Since the temporal appliance event positions are implicitly known from the consumer labeled time range, these event segments can be used to train a supervised event model for the event classification. The a priori known event segments can be used to identify significant event characteristics, which are a major advantage compared to hand-crafted rules. In a supervised classification task, the classifier needs training samples for each individual class. Event detection is related to anomaly detection that faces the problem of not having sufficient training samples for one of the targeting classes. In practice, we explicitly know from examples how an event looks like, but we don't explicitly know how a non-event looks like.

\begin{figure}[htbp]
\centering
\includegraphics[width=0.95\linewidth]{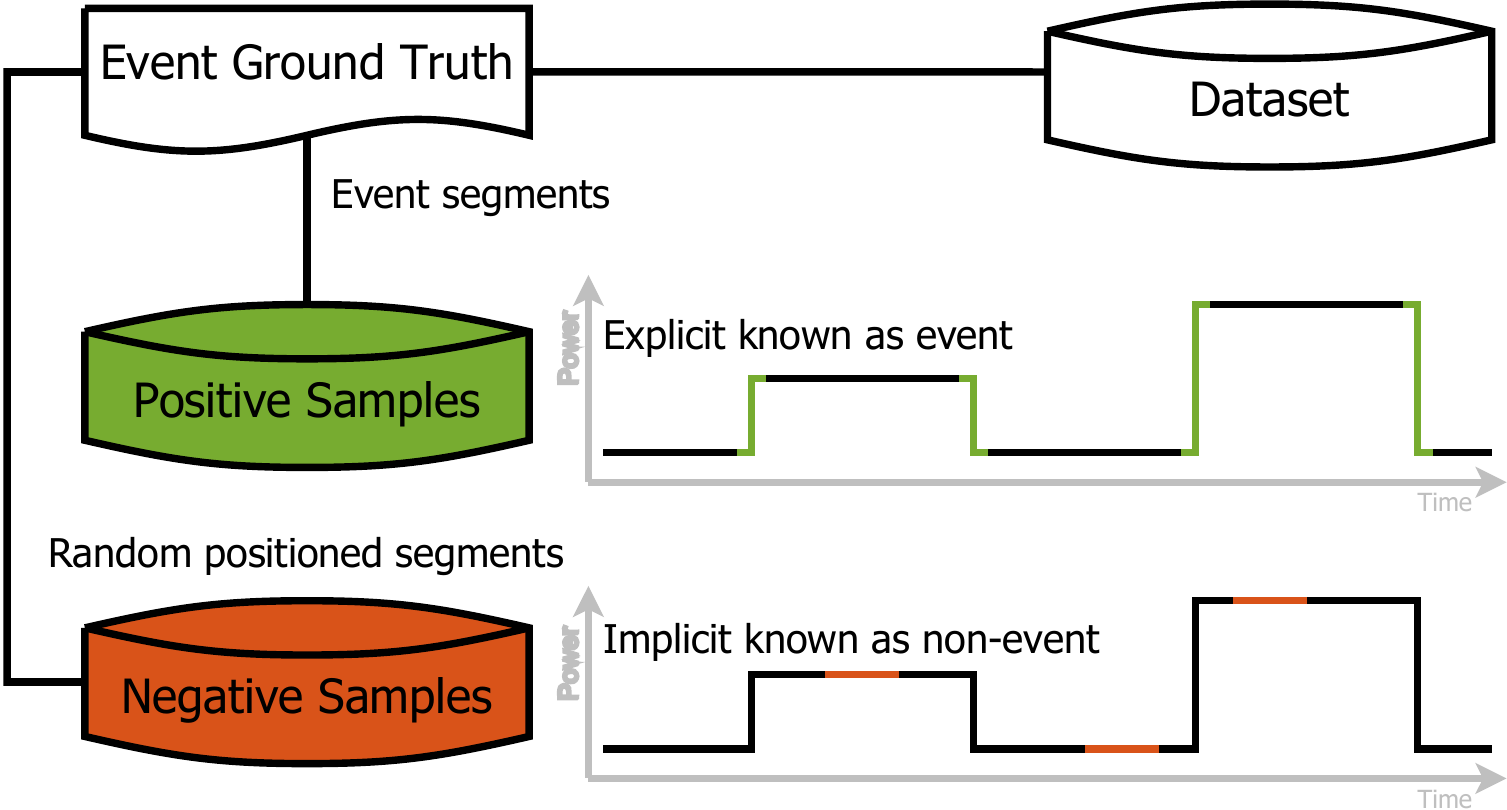}
\caption{The explicitly-known events are retrieved from the event ground truth. Therefore, all other regions are implicitly-known as non-events.}
\label{fig:ExplicitImplicitEvents}
\Description{Splitting the dataset segments in events and non-events using the event ground truth.}
\end{figure}

To overcome that issue, we make use of the fact, that statistically the majority of the time, no event occurs in the signals time domain. We cut short, randomly positioned regions of the temporal signal from the training area, to use them as non-event samples (see Figure~\ref{fig:ExplicitImplicitEvents}). The probability to hit an event on a randomly selected position in the training area of the temporal signal is low for common residential and office environments. Around 1\,250 events occur per phase in one week for the residential environment while it is around 257 for the office environment, based on the utilized datasets BLUED and BLOND-50. Assuming we are interested in the same number of non-events as it is for events, the chance to hit an event via random selection lies at 0.83\,\% for the residential environment, while it is around 0.17\,\% for the office environment. To even overcome that small uncertainty, a minimum temporal distance to explicitly known events of minimal 10\,s must be fulfilled. The resulting non-events will be named \emph{implicitly-known non-events} throughout this paper. All samples together can be used to train a classifier with a training set that consists of explicitly-known events and implicitly-known non-events.

An observed issue with this approach lies in a high number of event false positives. The randomly selected non-event samples stem mostly from areas of a steady consumption. Therefore the non-event class is a good homogeneous representation of steady non-event areas. A more heterogeneous set of non-event training samples with unsteady event-like transients would be necessary to improve the classification performance of transients from SMPS-driven appliances in favor to non-events.

\subsection{Adaptive Training}
Extracting even more randomly selected samples would be one infeasible way to get a higher variance. The extreme form would be to use every extractable time window in the dataset that is not a ground truth labeled event. Obviously, this would create an infeasible number of training samples for the non-event class. However, the vast amount of training samples would be unnecessary anyway due to a very strong similarity.

\begin{figure}[htbp]
\centering
\includegraphics[width=0.95\linewidth]{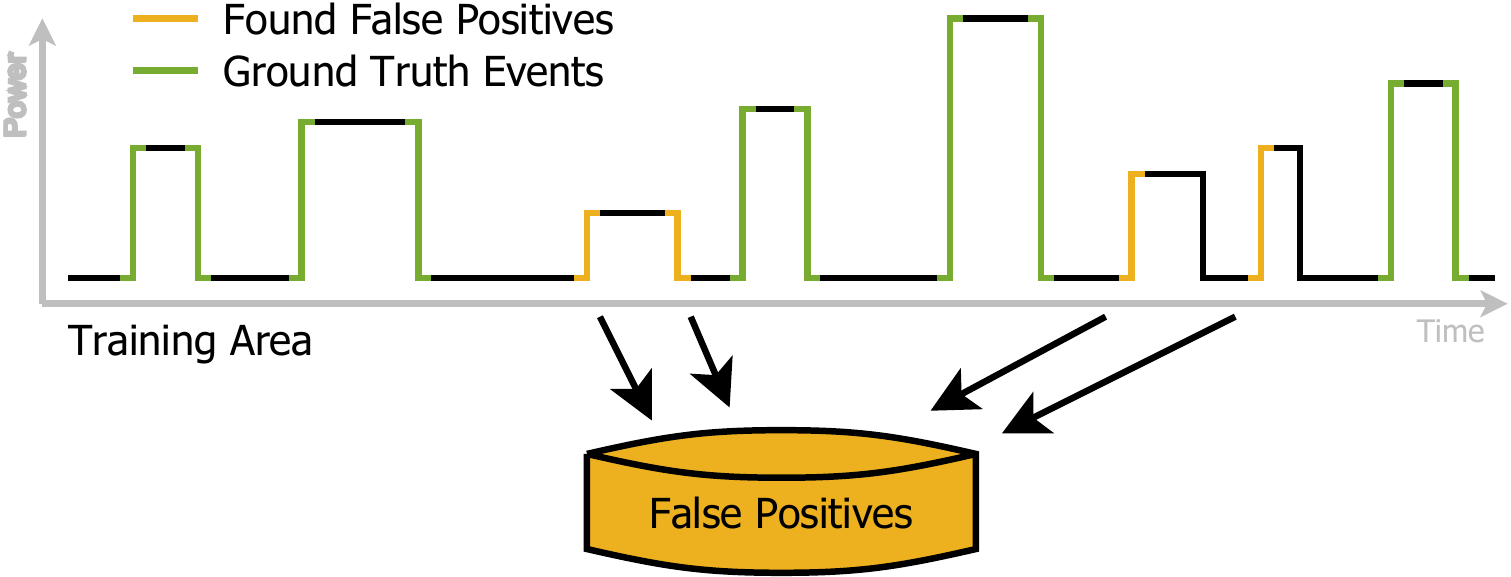}
\caption{The event detection runs on the training area and generates false positives that are being stored for the actual event detection.}
\label{fig:AdaptiveTraining}
\Description{Collecting False Positives from the training run.}
\end{figure}

Our approach is a so called boosting variant that runs the event detection algorithm on the whole training area to find all ground truth labeled events but also a certain amount of non-labeled transients. These transients are obvious false positives, based on the provided ground truth (see Figure~\ref{fig:AdaptiveTraining}). They are marginal, uncertain segments of non-events that share similarities with events. These similarities cause the misclassification in favor to the class \emph{event}. Since these false positives are found inside the training set, we can use them freely to improve our classification model.

\begin{figure}[htbp]
\centering
\includegraphics[width=0.95\linewidth]{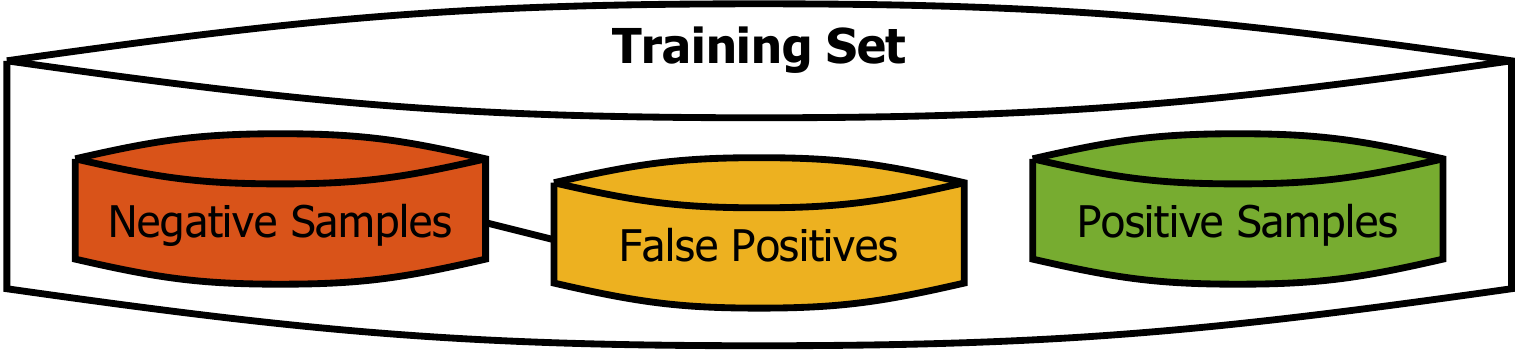}
\caption{The collected false positives from the event detection of the training area form together with the negative samples the class non-events. The positive examples are the representatives of the event class.}
\label{fig:MultiVarAdaptEventDetection}
\Description{Contents of the training set: positive samples and Negative samples together with false positives.}
\end{figure}

The idea lies in adding these \emph{edgy} transients to the non-events class of the training set to improve the border between events and non-events (see Figure~\ref{fig:MultiVarAdaptEventDetection}). The actual training set consists now of ground truth labeled event samples, implicitly-known non-event samples, and false positives that were found in the event classification run on the training area itself. This way, it is possible to overcome the issue of finding proper non-event samples for the event detection algorithm. To even reduce the amount of false positives further, the adaptive training can be applied multiple times.

\subsection{Event Features}
The event ground truth information for BLUED is based on a power consumption change of at least 30\,W over a time period of minimal 5\,s \cite{Anderson2012}. Based on this definition, the appliance events can be identified in a moving time window in the continuous electricity signals. We implemented one spectral and six time domain metrics as appliance event features for the classification between events and non-events. Our design defines that the actual event transient is being aligned in the middle of the extracted time window with 5\,s of data before and after the actual event transient. The actual temporal position of the event transient is being extracted from the ground truth information or manual annotation in the case of BLOND-50.


The BLUED provided ground truth information and BLOND-50 annotations from this work, comprises the appliance ON and OFF switch events, including circuit number, temporal position (timestamp) and appliance type. The provided switch-OFF and switch-ON events of these appliances will always cause significant changes in these consumption-related metrics:

\subsubsection*{Current}
The current is the first intuitive metric that contains consumption changes (see Figure~\ref{fig:eventFeatures}-1). The RMS current $I_{rms}$ for each period is calculated as follows, with $N$ as the number of samples per period, calculated as the ratio of the sampling frequency $fs$ and the mains frequency $F0$.

$$ I_{rms}(p) = \sqrt{ \frac{1}{N}\sum_{k=1}^{N}{I_k^2} }, ~~~ N=fs/F0 $$
$$ \vec{I}_{rms} = [I_{rms}(1), I_{rms}(2), \dots I_{rms}(nPeriods)] $$

\subsubsection*{$\Delta$(Current)}
Since multiple appliances can run at the same time, the actual pre-event current can be a sum of multiple appliances and therefore has a high variance (see Figure \ref{fig:eventFeatures}). The actual information of interest is the current step change at the event time (see Figure~\ref{fig:eventFeatures}-2). This metric can be retrieved by the numerical difference of the neighboring elements of the current periods $\vec{I}_{rms}$. The operation is the derivation equivalent for discrete time series.

$$ \Delta \vec{I}_{rms} = \vec{I}_{rms_k} - \vec{I}_{rms_{k+1}} $$
$$ \vec{I}_{rms_k} = [I_{rms}(1), \dots, I_{rms}(k-1)] $$
$$ \vec{I}_{rms_{k+1}} = [I_{rms}(2), \dots, I_{rms}(k)] $$

\subsubsection*{Admittance}
The grids voltage can contain high fluctuations (up to 10\,\%), which influences the current signal as well. The admittance removes the voltage influence from the current signal and is therefore more precise to the appliance consumption itself (see Figure~\ref{fig:eventFeatures}-3). The admittance ADM, can be calculated by the element wise vector division of the period wise current $\vec{I}_{rms}$ and voltage $\vec{U}_{rms}$.



\subsubsection*{Spectral Flatness}
Our motivation for the only spectral feature we considered is the assumption that all appliances have their individual fingerprints in their harmonic energy distribution. A suitable spectral one-dimensional metric is the spectral flatness. A flat spectral curve $f_{bins}$ would cause a value close to one, while a single strong spike would lead to a value close to zero (see Figure~\ref{fig:eventFeatures}-4). The switch-OFF and switch-ON of an appliance influences the spectral flatness in general way. The spectral flatness $SPF(p)$ for each period is calculated by the ratio of the geometric and the arithmetic mean of the current signal energy spectrum \cite{Peeters2004}.

$$SPF(p)=\frac{{\sqrt[ {N}]{\prod_{f \in f_{bins}} x_f}}} {\frac{1}{N} \sum_{f \in f_{bins}} x_f}$$

\subsubsection*{Cumulative Sum}
The cumulative sum is a sequence analysis technique that allows to identify small and continuously slow as well as strong and fast changes in a sequential time series (see Figure~\ref{fig:eventFeatures}-5). It is therefore a common technique for change and event detection. The cumulative sum is the sum of the differences to the mean of the signal in between a defined time window.

\subsubsection*{$\Delta$(Cumulative Sum)}
\begin{figure}[htbp]
\centering
{\includegraphics[width=0.495\linewidth]{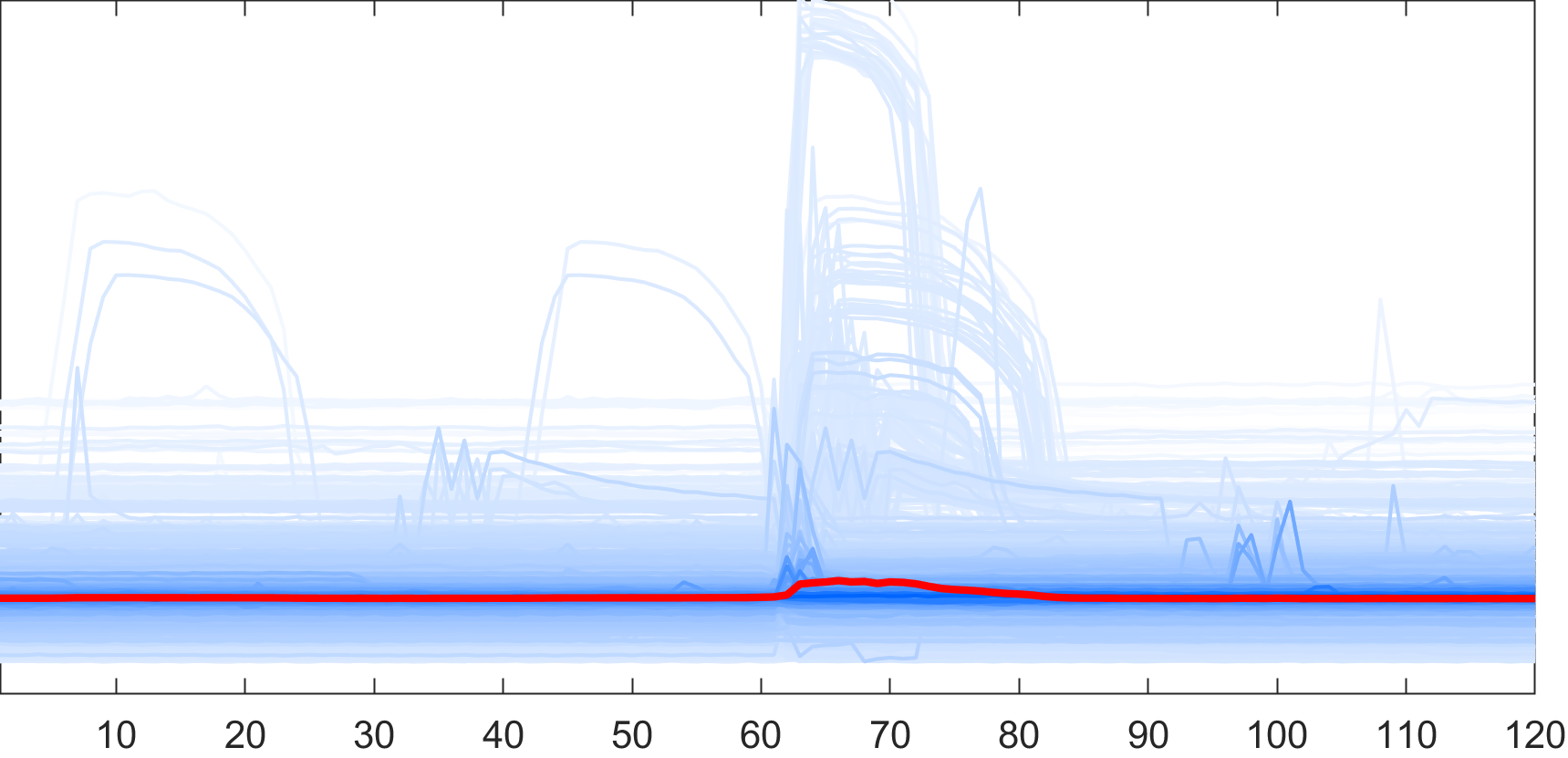}}
{\includegraphics[width=0.495\linewidth]{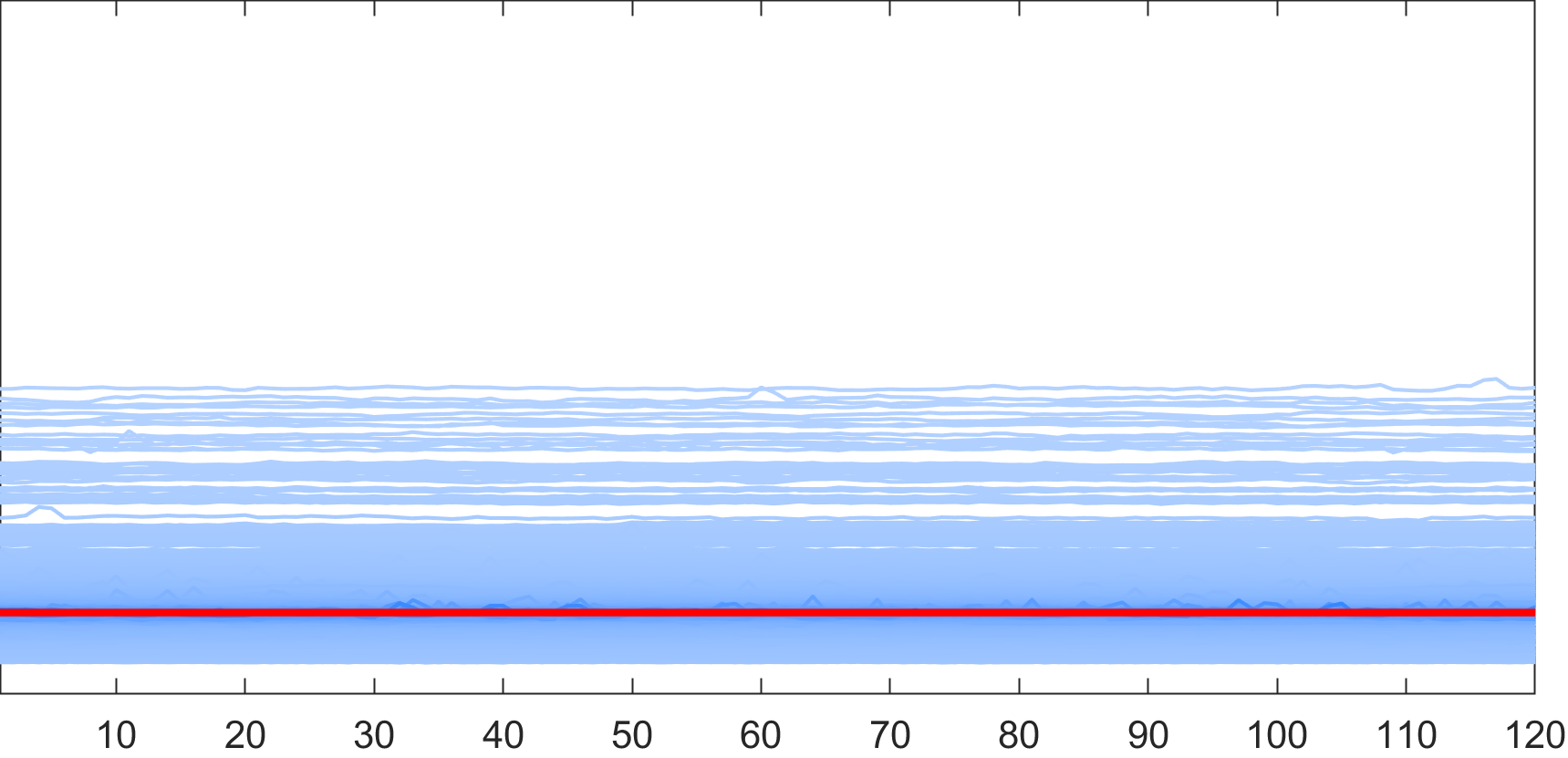}}
{\includegraphics[width=0.495\linewidth]{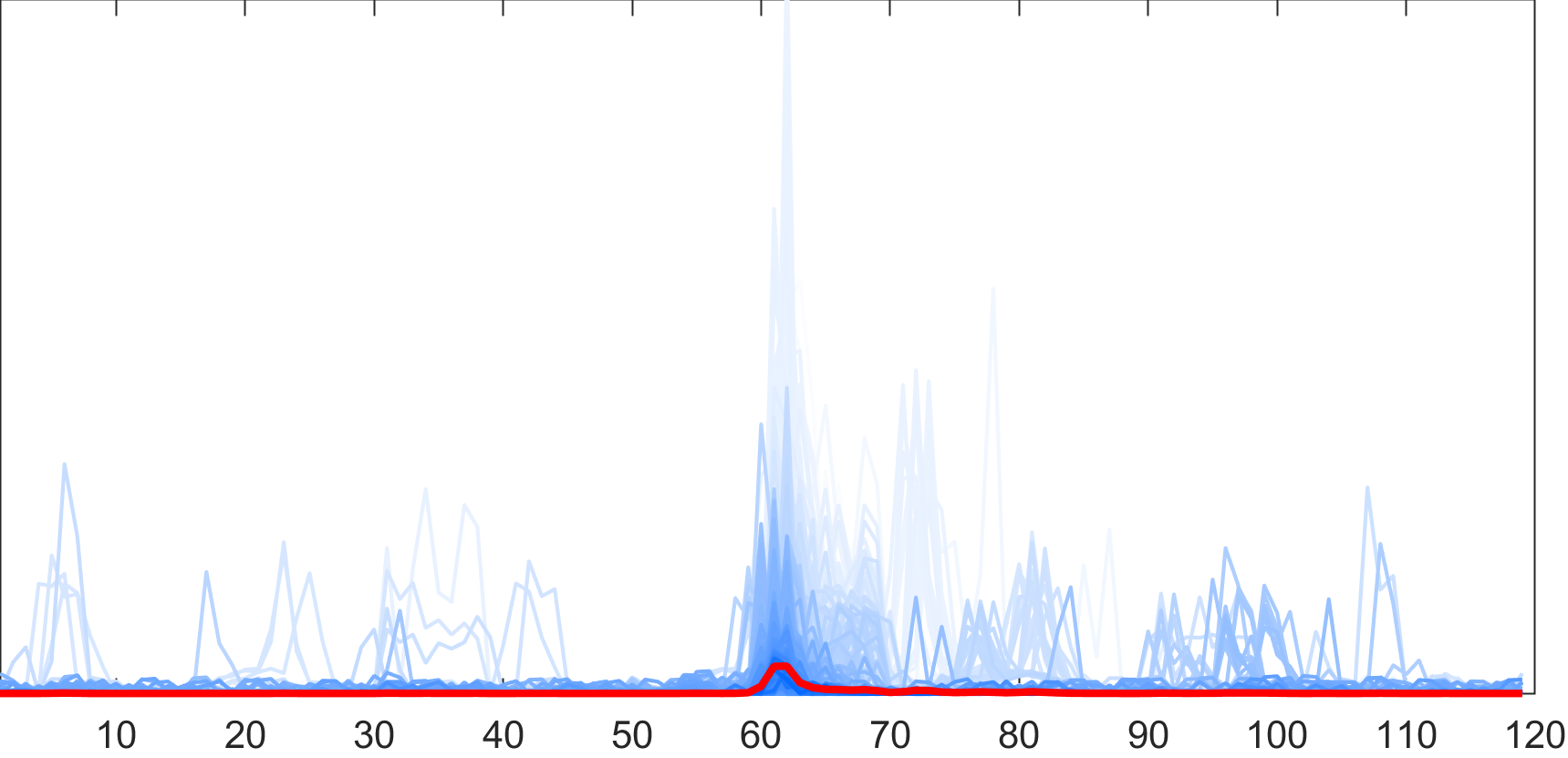}}
{\includegraphics[width=0.495\linewidth]{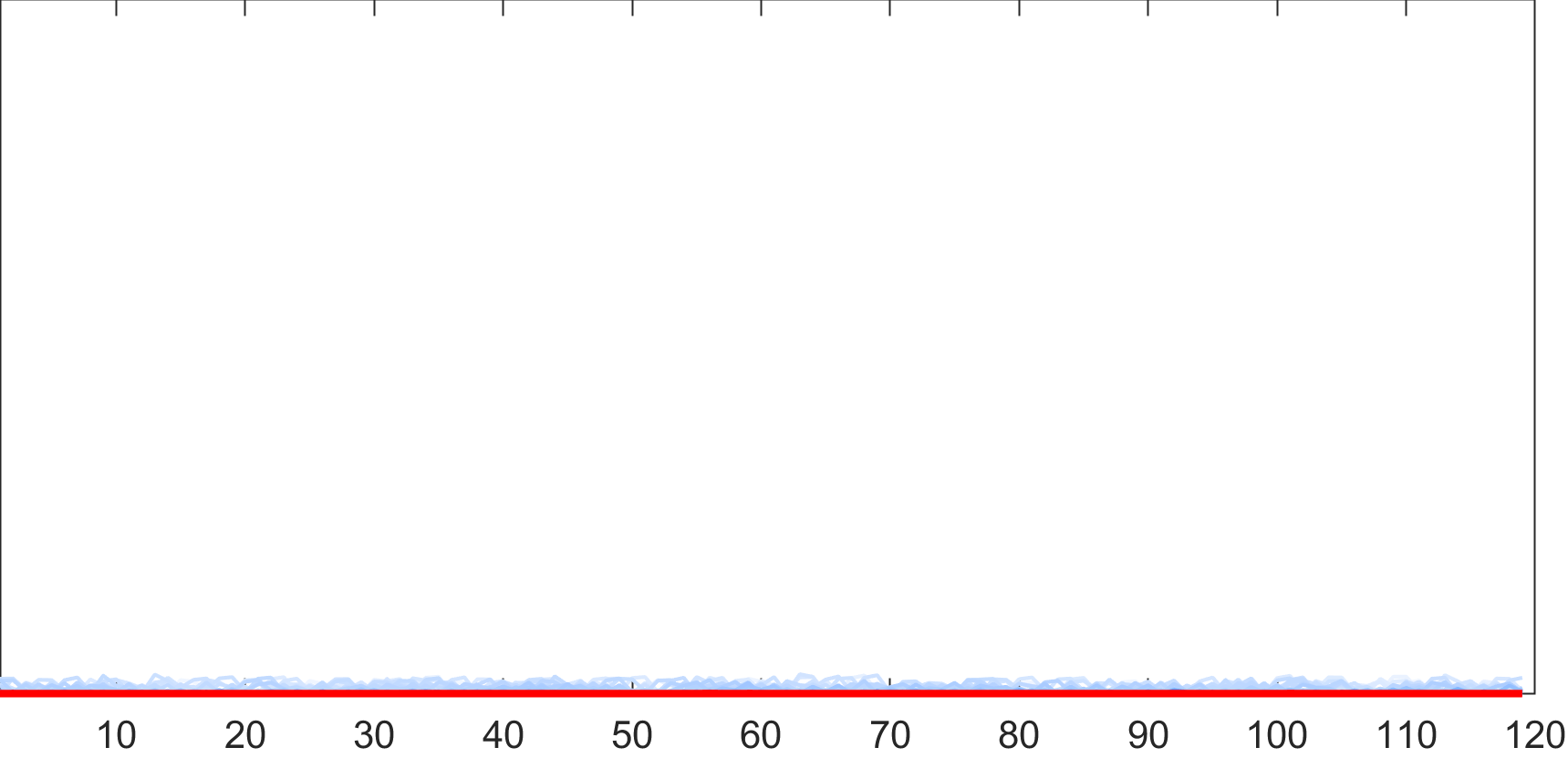}}
{\includegraphics[width=0.495\linewidth]{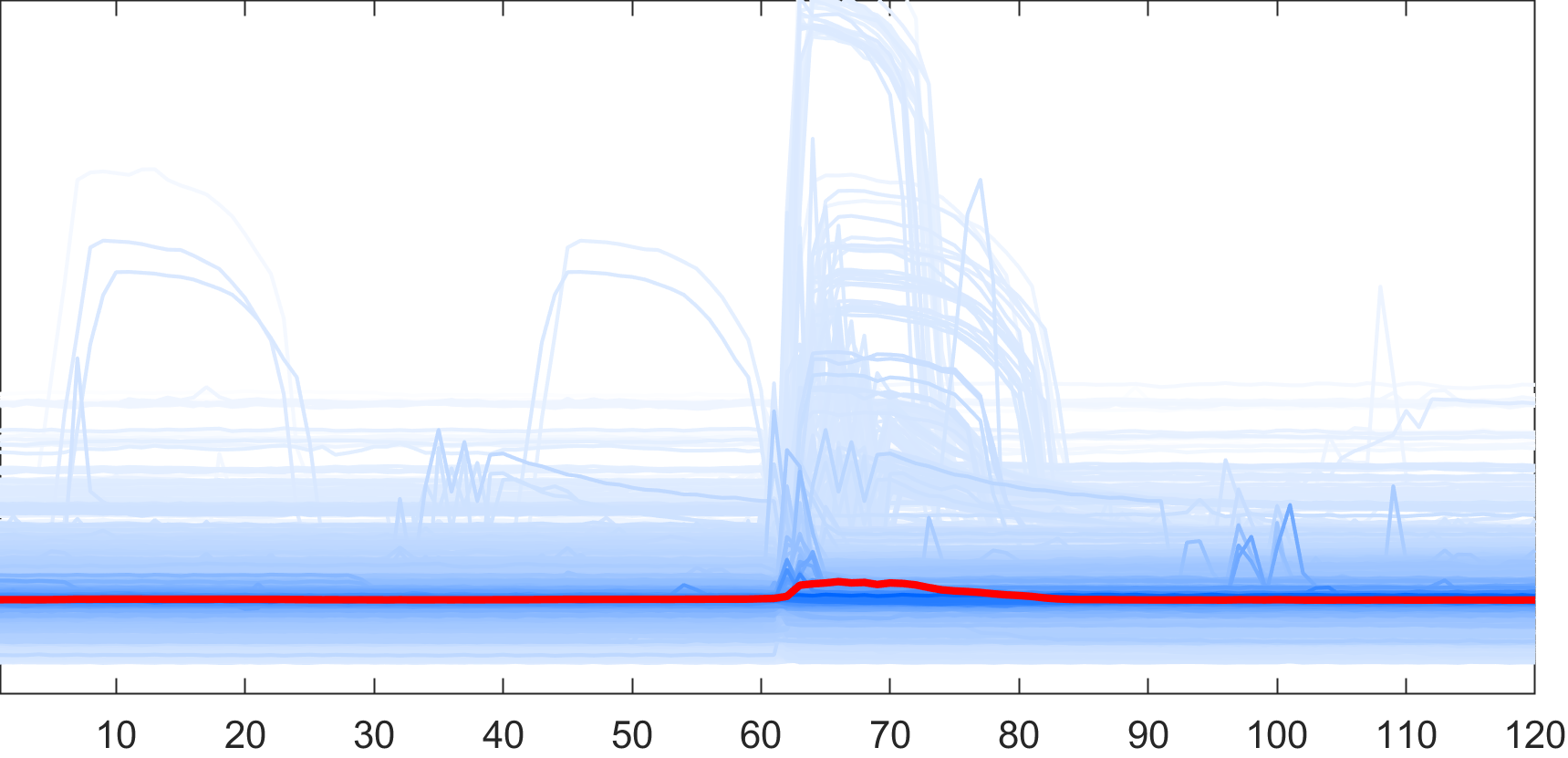}}
{\includegraphics[width=0.495\linewidth]{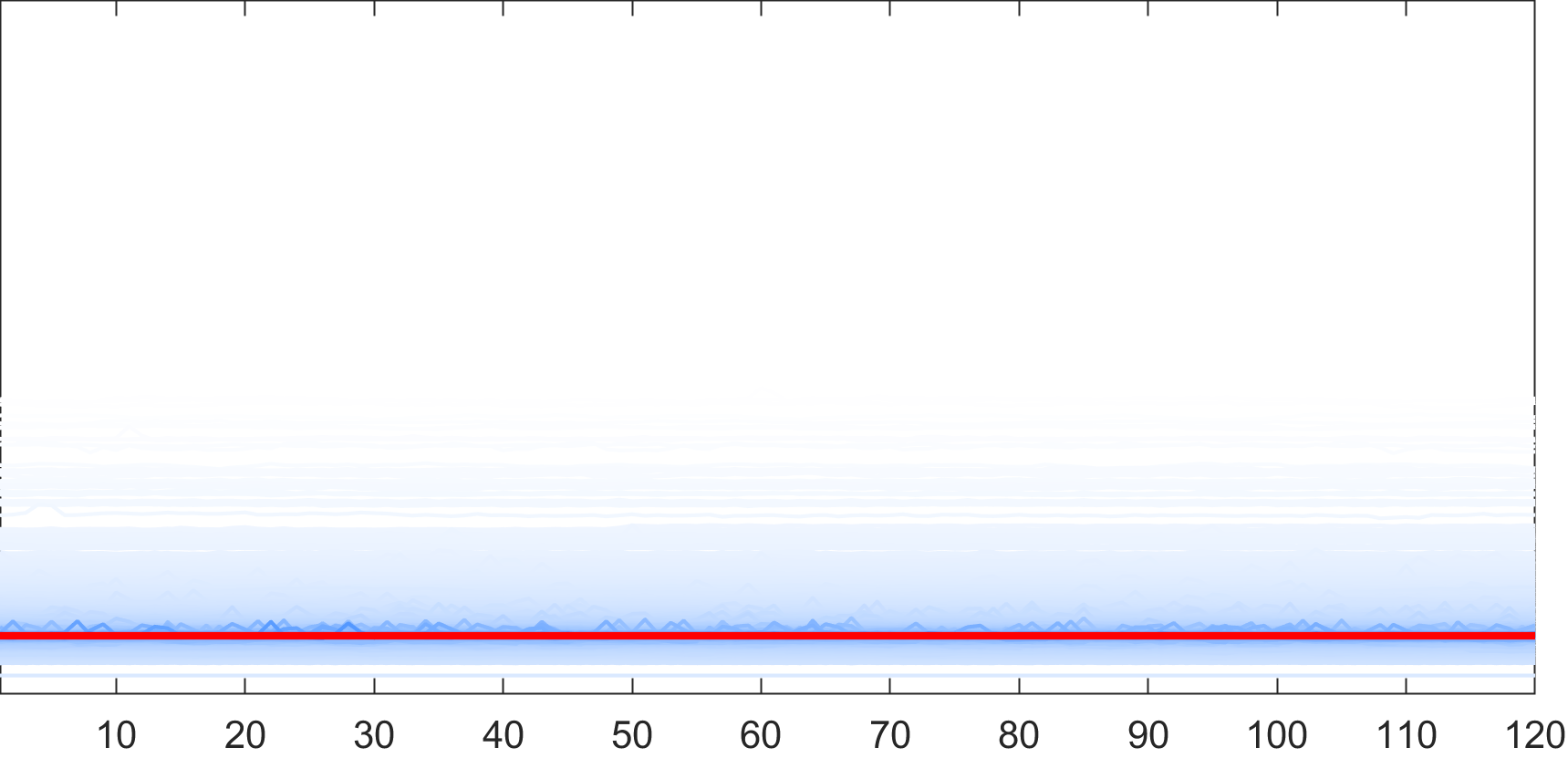}}
{\includegraphics[width=0.495\linewidth]{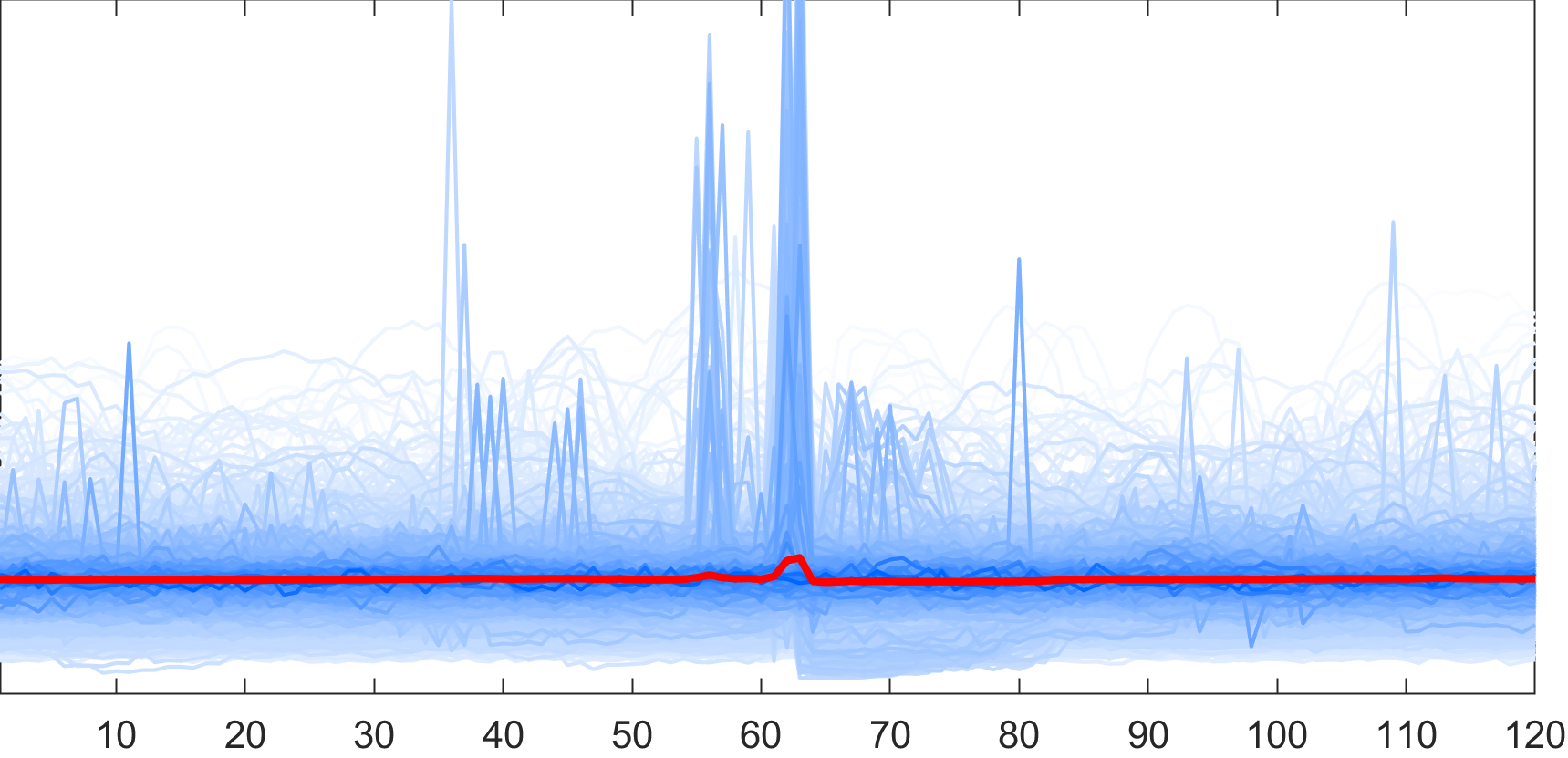}}
{\includegraphics[width=0.495\linewidth]{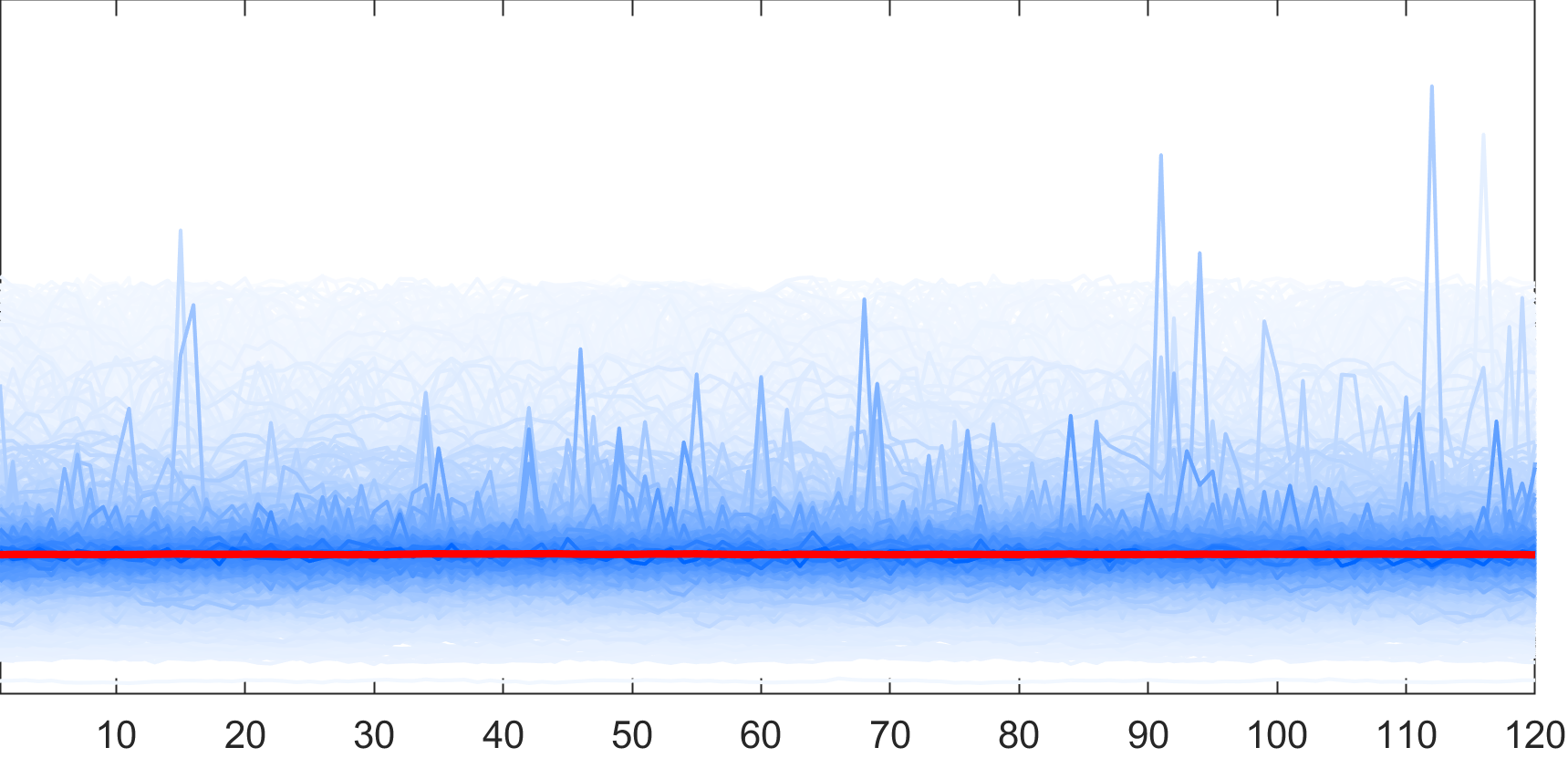}}
{\includegraphics[width=0.495\linewidth]{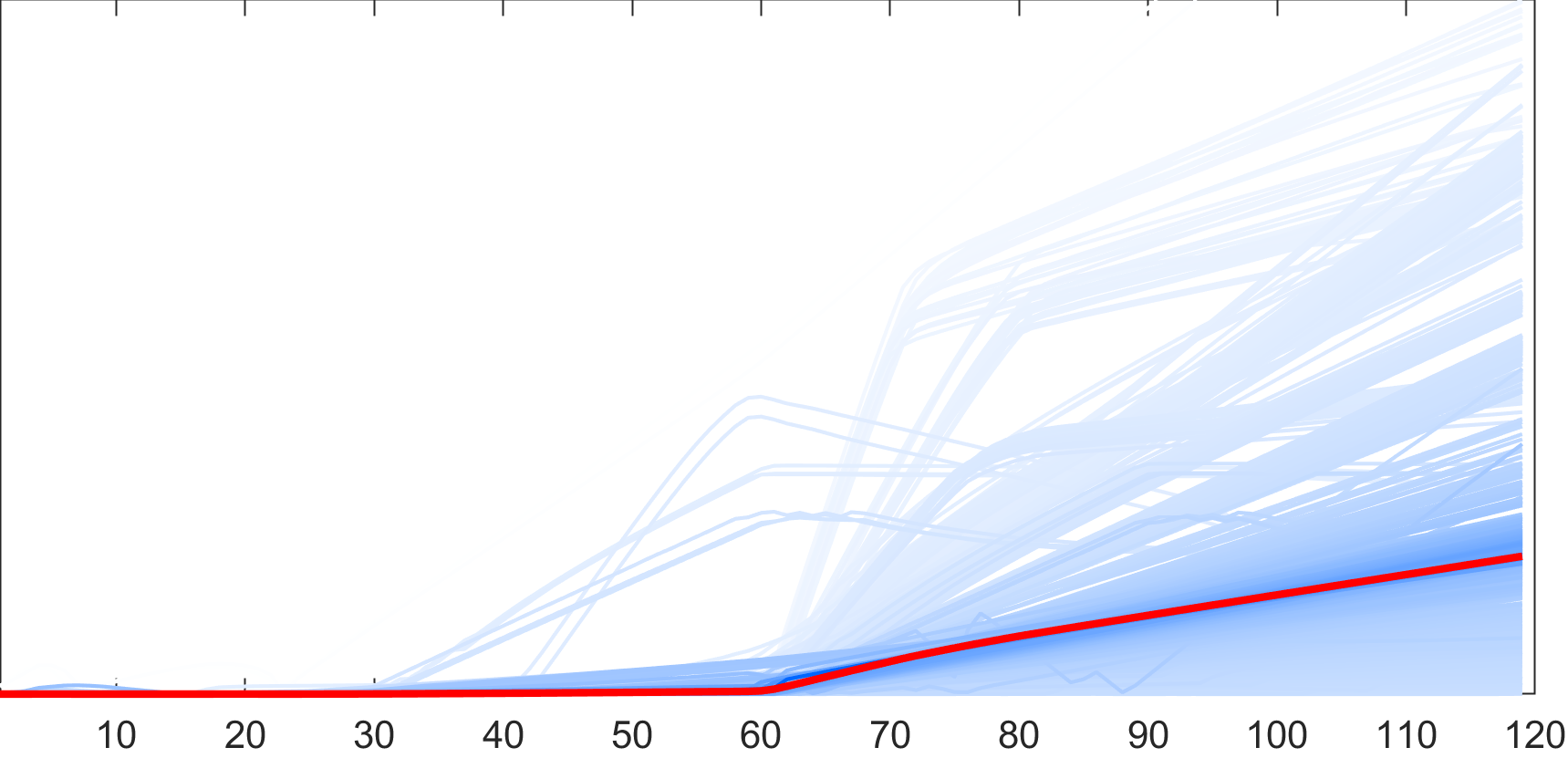}}
{\includegraphics[width=0.495\linewidth]{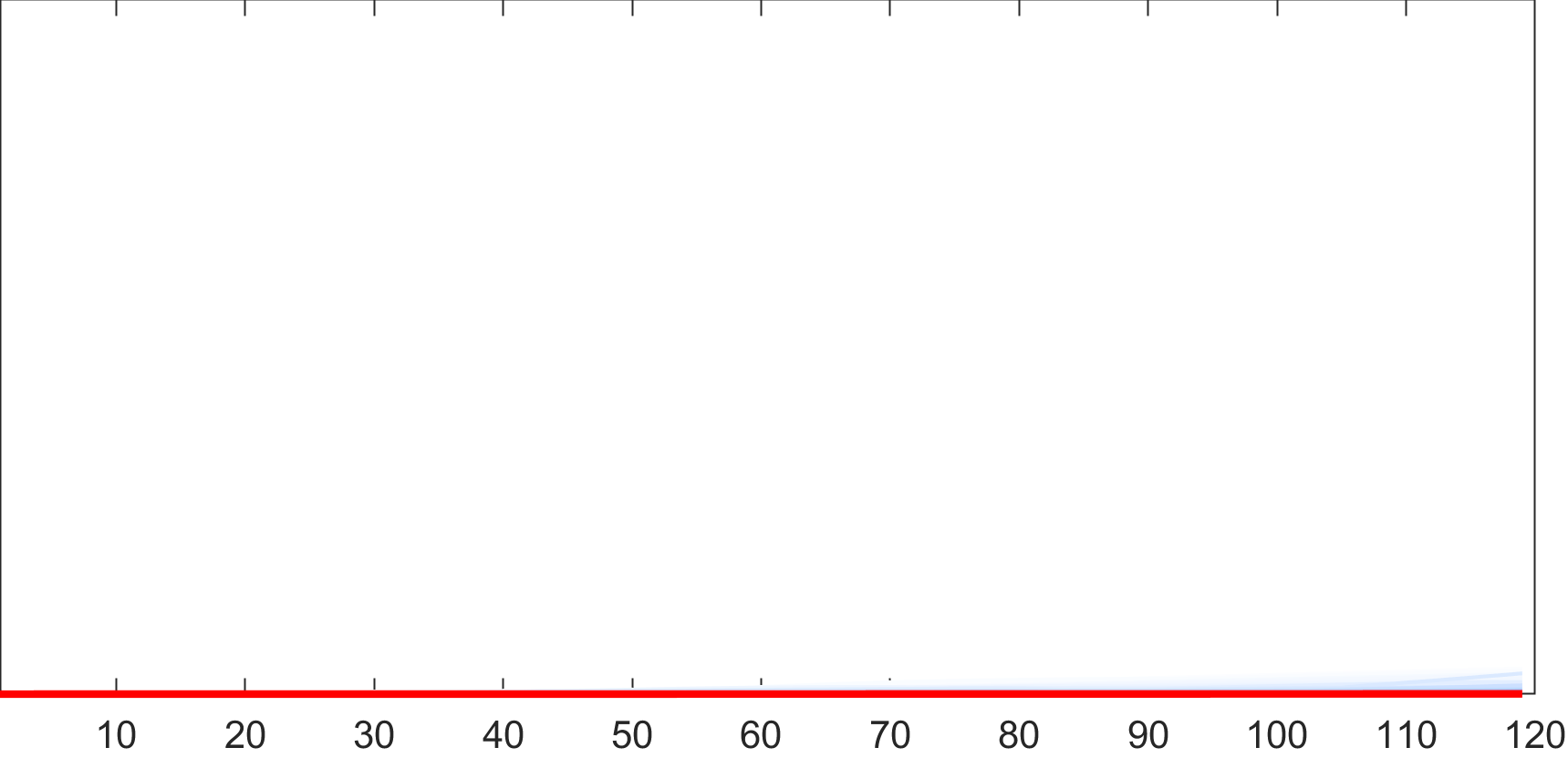}}
{\includegraphics[width=0.495\linewidth]{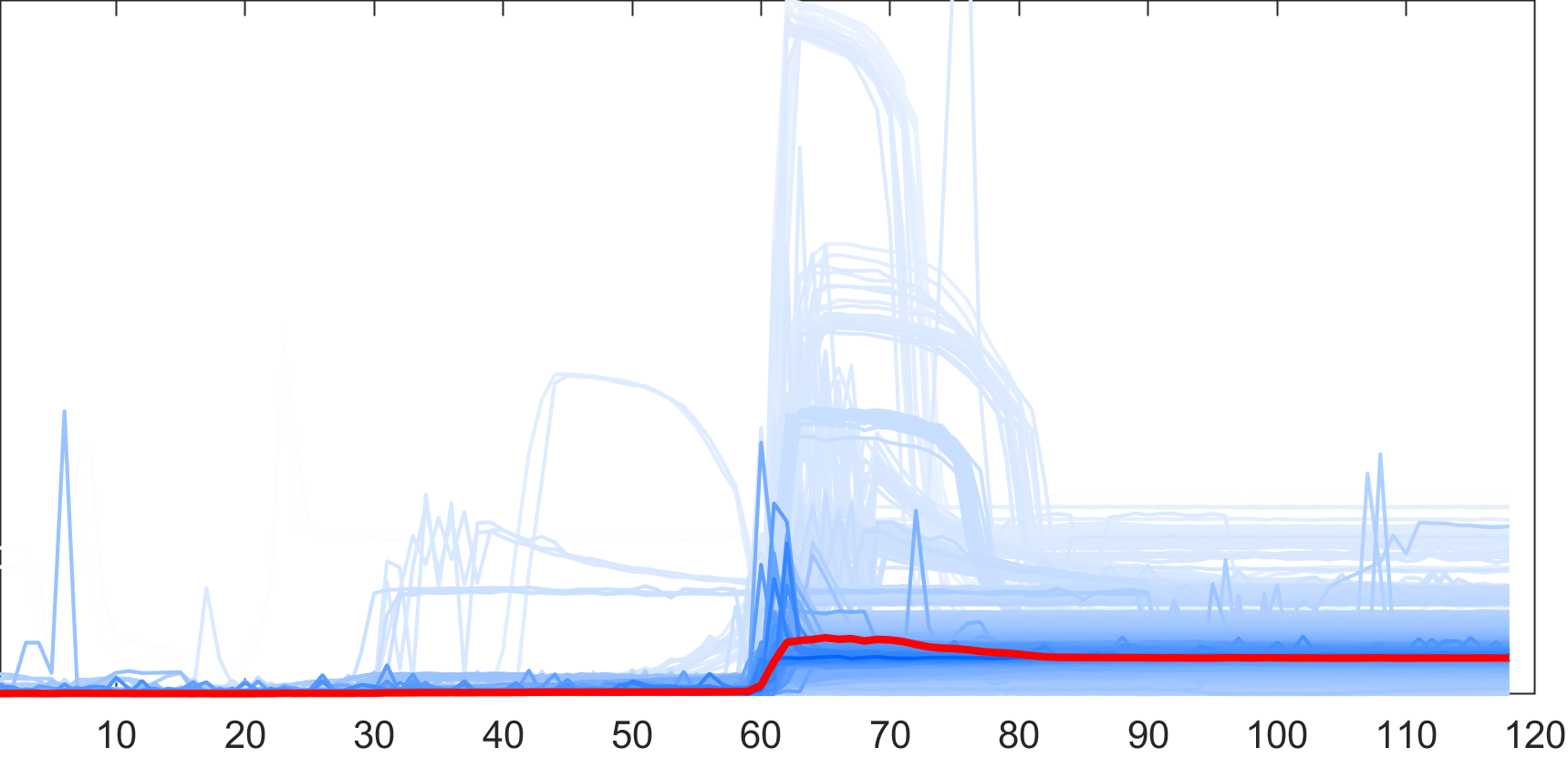}}
{\includegraphics[width=0.495\linewidth]{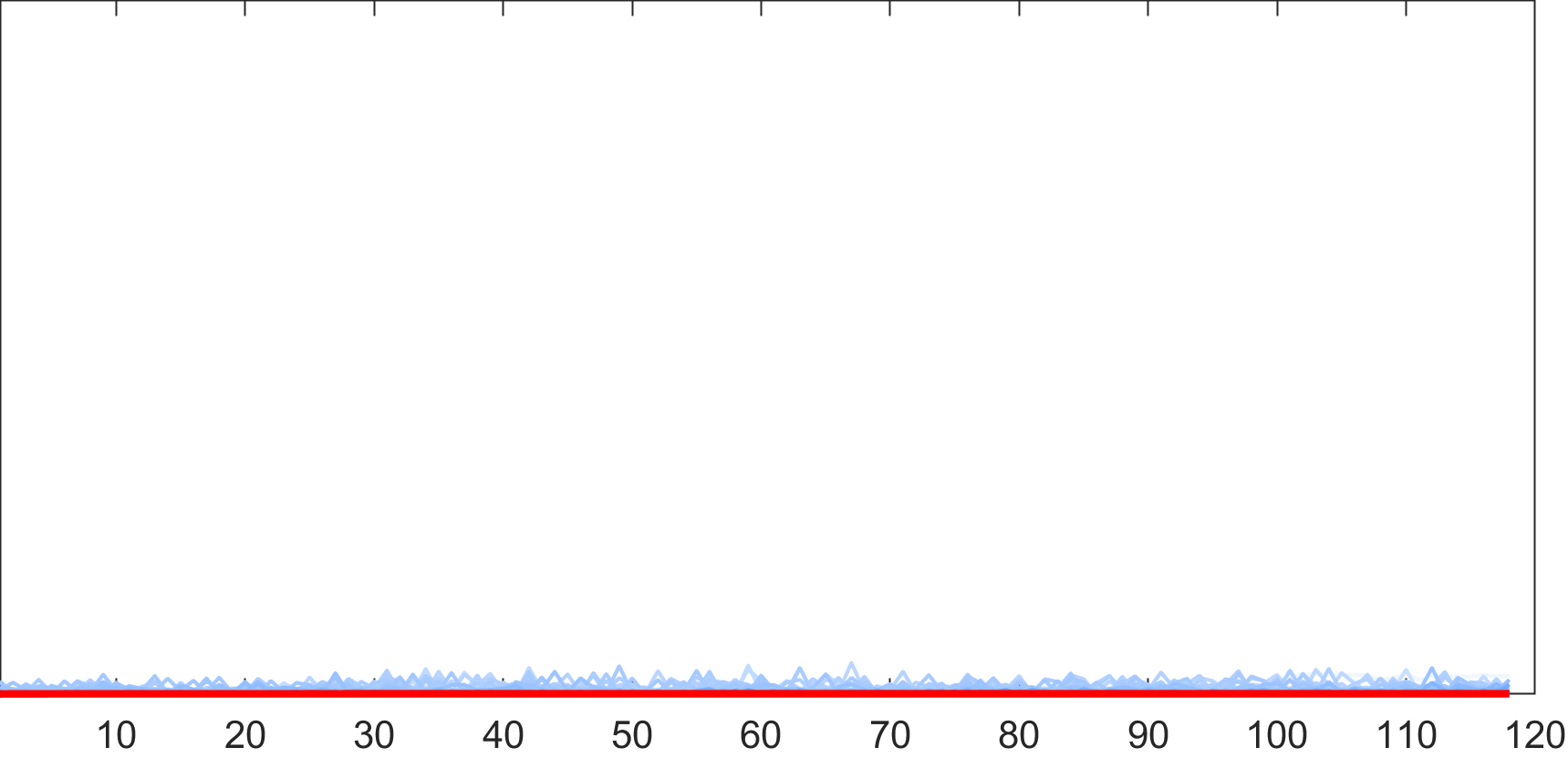}}
\caption{Events (left) and Non-Events (right) with Periods in the X-Axis and amplitude in the Y-Axis of the 6 event feature metrics: 1. Current, 2. $\Delta$(Current), 3. Admittance, 4. Spectral Flatness, 5. CUSUM and 6. $\Delta$(CUSUM). The color saturation correlates with the average distance to the mean event (red line). The closer the event lies to the mean-event the higher is the saturation.}
\label{fig:eventFeatures}
\Description{event feature metrics plotted for each event and non event.}
\end{figure}

The cumulative sum can have extreme gains in their values and therefore causing undesired weighting of dimensions in the feature space. The derivative of the cumulative sum is a way to prevent this issue and to keep the values in a lower magnitude. The resulting signal is visually comparable with the current itself, but with enlarged transients (see Figure~\ref{fig:eventFeatures}-1 and \ref{fig:eventFeatures}-6).

$$ \Delta \vec{I}_{cms} = \vec{I}_{cms_k} - \vec{I}_{cms_{k+1}} $$

In addition to the mentioned features and training methods, we evaluated the event detection performance through different methods in the feature space normalization and classification step. To avoid undesired weighting across the dimensions of the feature space, a common technique is to apply a feature space normalization. This is often an essential step, of which we evaluate three types. The classification step is being evaluated with two different classifier (KNN and SVM) including their hyper-parameter search.


\section{Experiments}

To compare our event detection performance with state-of-the-art, we applied our algorithm on the BLUED dataset, which is commonly used for event detection evaluation. The experimental setup is oriented on the setup in the work of \citet{DeBaets2016}. While De Baets is using a fixed test area, we are using cross-validation for our performance evaluation. At least \citet{Anderson2012a}, \citet{Barsim2014} and \citet{Wild2015} evaluate their event detection algorithm on the BLUED dataset as well. For BLUED we use the provided ground truth information which stems from hand-crafted annotations.

Unfortunately, neither BLUED nor BLOND-50 provide versatile event information that allows a determination between ON / OFF-switching and user-unrelated transients. In our experiments on BLOND-50, we try to distinguish ON and OFF events from all remaining state transients - identical to the work of \citet{DeBaets2016}. The appliance ON and OFF events for the BLOND-50 dataset are being collected by visual observation of an instructed person with the help of a self-implemented annotation tool. There are no studies regarding event detection on BOND-50 yet.

Since the benchmark of several parameters using cross-validation takes much computational time, we use a cluster of 60 virtual machines, based on dual Intel Xeon E5-2630v3 with each four cores and 10\,GiB RAM to execute the appliance event detection algorithm in parallel. The cumulative CPU time for all experiments, preprocessing and testing lies in a range of 128\,000 CPU-core-hours.

\begin{figure}[htbp]
\centering
\includegraphics[width=0.95\linewidth]{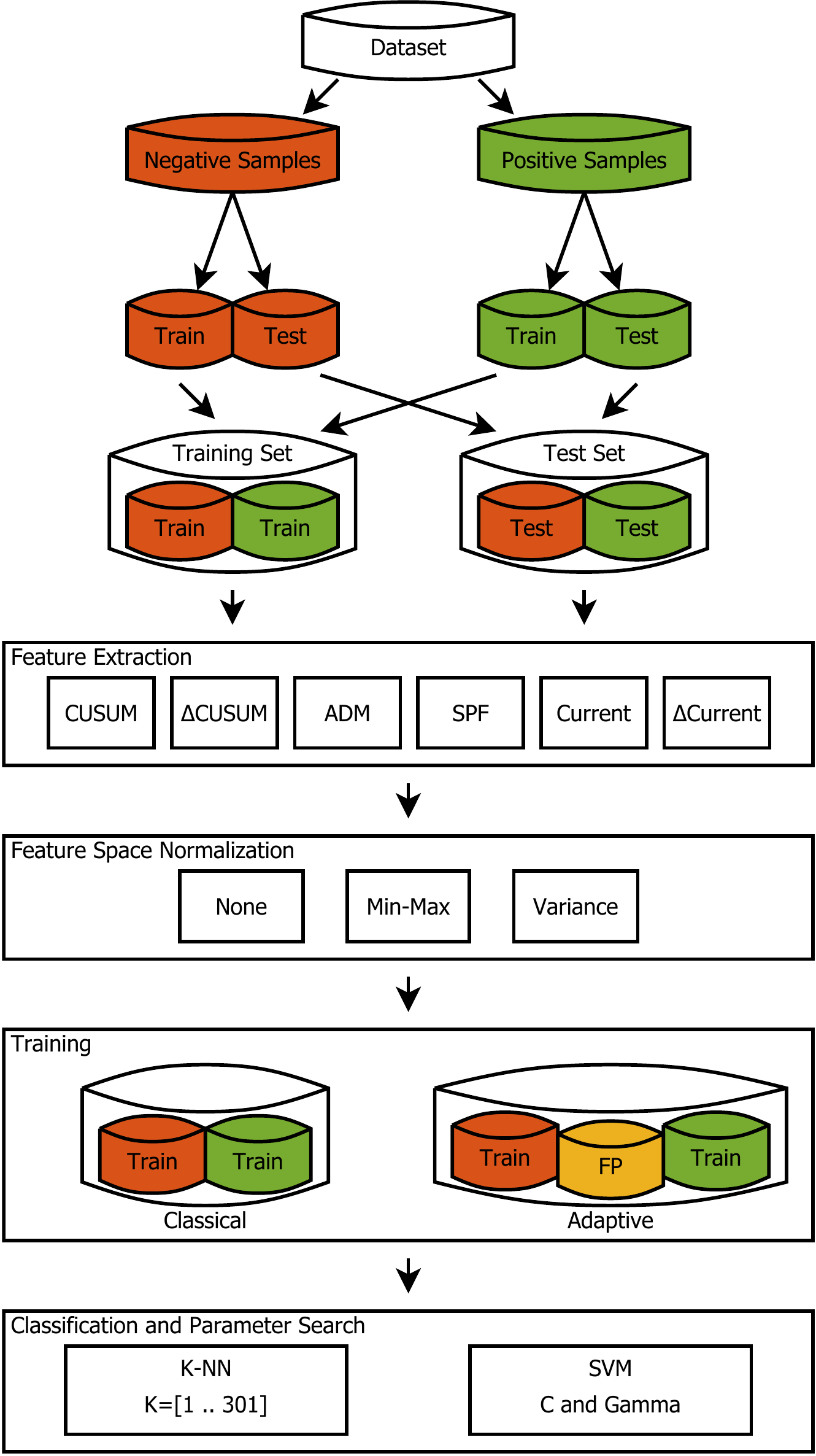}
\caption{The architecture of the experimental setup considers the main step of the common machine learning pipeline and includes the evaluation of six event features, three types of feature space normalization, two training approaches and two different classifiers with its optimal parameters. The whole architecture is wrapped by a cross-validation and structured to run on a distributed computation system.}
\label{fig:experimentsOverview}
\Description{The architecture of the experimental setup.}
\end{figure}

\subsection{Multivariate Event Detection}
Instead of monitoring one or few parameters passing thresholds, our multivariate approach enables supervised learning of multiple event characteristics. The explicitly-known event, and implicitly-known non-event sections were used to train the classifier that decides, based on the given feature vector, between event or non-event.

\paragraph{Architecture for BLUED}
In addition to the 1\,577 events, we extracted 6\,428 segments of implicitly-known non-events (one for each file) of the same length. The segments are aligned with the ground truth event timestamp in the center. These segments are fed to the feature extraction and normalization after that. The normalization parameters (e.g., means or standard deviation) are saved to apply the corresponding transformation to the samples of the test area. The following steps include a parameter search for the classifier (e.g., C and Gamma for SVM), classifier training, and classification of the samples of the test area (see Figure~\ref{fig:experimentsOverview}). All experiments are implemented within a stratified k-fold cross-validation to ensure reliable results.

\paragraph{Architecture for BLOND-50}
The manually annotated temporal time span comprises one month of measured data. We extract all manually annotated events and the implicitly-known non-events in a very similar way as we do for BLUED. This step yields in 3\,310 event and 3\,264 non-event samples. The events originate from 41 different monitored appliances in the time range of \texttt{2016-11-01} to \texttt{2016-11-30}. 

\subsection{Adaptive Training}
The adaptive training shares the same experimental architecture as the multivariate event detection, with one additional event detection run on the training area itself and its false positives included to the training set. This training run finds events in the training area that can be divided, considering the ground truth information, into true positives, false positives, and false negatives. All false positive segments that originate from the training run are added to the non-event class of the actual training set. 

\subsection{Manual BLOND-50 Event Annotation}
Every performance benchmark needs reference information to enable comparisons. For the event detection evaluation, an event ground truth including the exact temporal position of an appliance event is necessary. For the BLUED dataset, the appliance events are provided already, for BLOND-50, the appliance events and the corresponding measurement system, circuit and socket number need to be acquired.

\begin{figure}[htbp]
\centering
\includegraphics[width=0.95\linewidth]{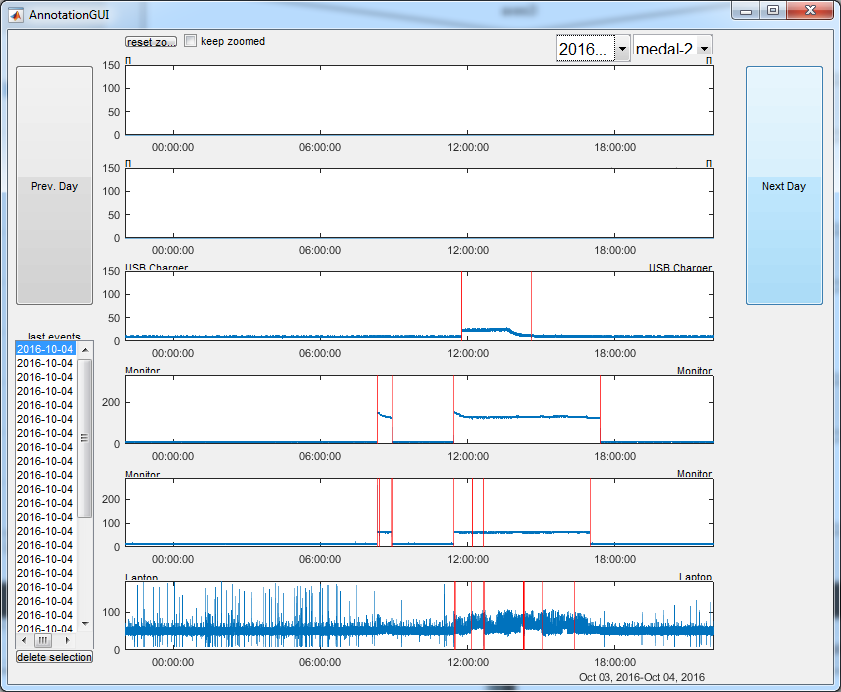}
\caption{Annotation tool for BLOND-50 event ground truth annotation. The annotating person specifies the date and measurement system to view the corresponding consumption of the day for each of the six sockets. Zooming into the time series plot allows for a precise event annotation.}
\label{fig:AnnotationTool}
\Description{Screenshot of the annotation tool.}
\end{figure}

To label the data with a ground truth, we were using a self-designed annotation tool that allows a manual annotation in the per-appliance subset of BLOND-50 (see Figure~\ref{fig:AnnotationTool}). The annotating person observes the data of one measurement system instance and all 6 sockets for one day per screen. The two appliance event constraints (power-rise/fall of 30\,W for a minimum time span of 5\,s) are communicated to the annotating person to ensure consistent events. In addition, the annotating person is instructed to consider only obvious appliance ON and OFF events. Transients that fulfill event constraints but are not obvious switch ON and OFF events are ignored.

The event ground truth for BLUED and BLOND-50 originate from visual time series observation by humans. Therefore the experimental evaluations in this paper are not performed on the (non-existing) absolute truth but rather subjectively chosen time series segments of the human observation that always contain an individual degree of uncertainty. Since neither an event ground truth nor an appliance event definition has been chosen, the goal is to retrieve an appliance event model from user chosen examples declaratively, a degree of uncertainty from the human observation therefore does not play any role. The manually annotated events, as well as the corresponding annotation tool for MATLAB, can be downloaded at the following link\footnote{Due to the double-blind review, the link will be available in the camera-ready version or can be requested via program chair.}.


\section{Results}
\begin{figure}[htbp]
\centering
\includegraphics[width=\linewidth]{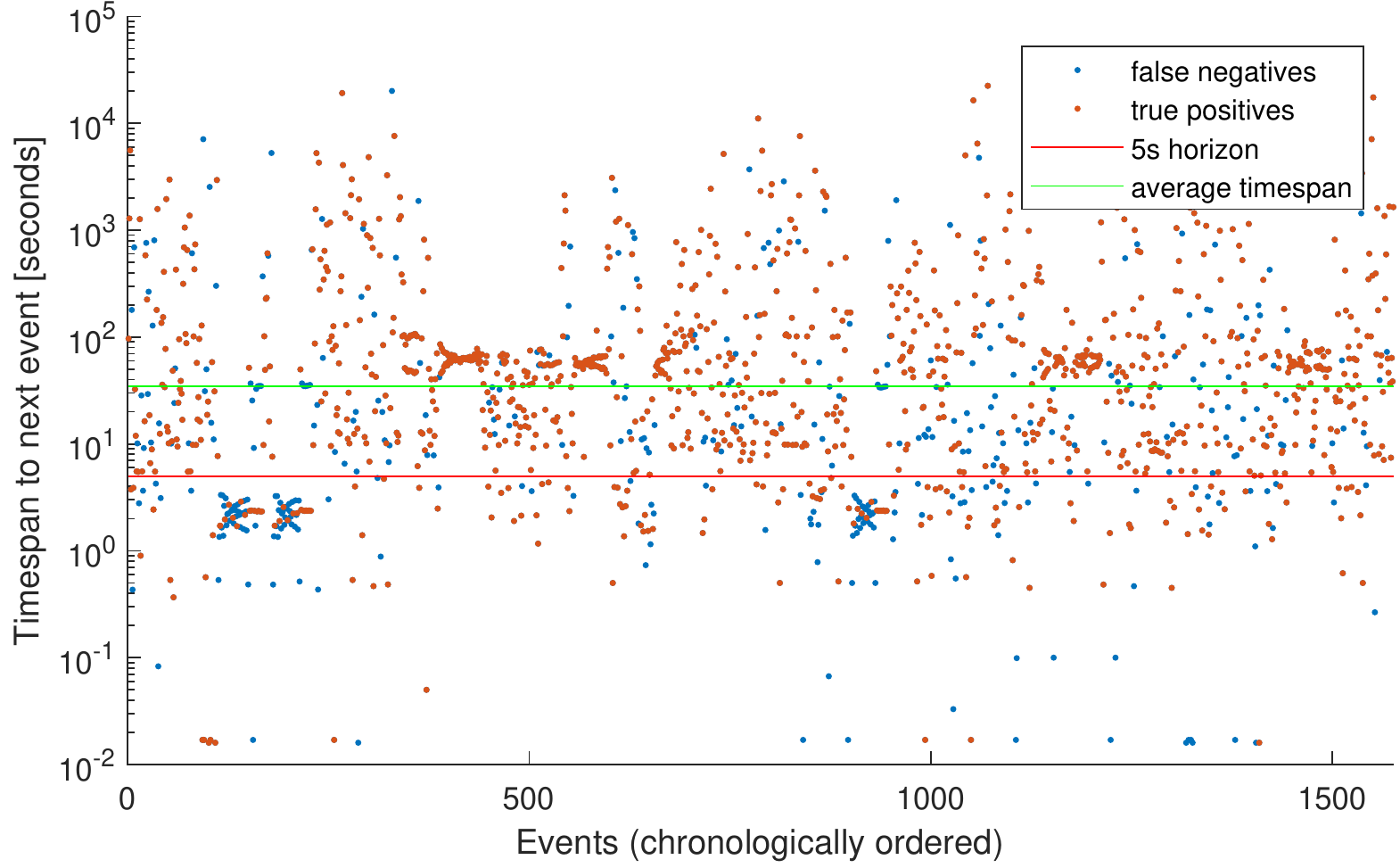}
\caption{A BLUED scatter plot of chronologically ordered events with their distance to the next event in the position of the y-axis. The three star-shaped clusters below the 5s horizon are caused by the printer appliance. The detection rate drops below the 5\,s horizon, causing more false negatives, due to the fact that 2 events in between 5\,s are recognized as one event.}
\label{fig:EventTimeSpanScatter}
\Description{Scatter plot of chronologically ordered events.}
\end{figure}

To ensure a consistent evaluation pipeline we decided to use the best parameter or settings from the previous steps. In practice, the evaluation of the normalization method is done with the best performing feature of the feature evaluation. For all experiments, a search window step-size of 30 periods was used. To the nature of the algorithms, multiple events occurring in between a 5\,s window (SCP violation) may be recognized as one event (see Figure~\ref{fig:EventTimeSpanScatter}). That circumstance causes a small number of false positives. The goal of all experiments is to find all ground truth labeled events (true positives) while keeping the misclassifications (false positives) to a minimum.

Precision, Recall, and F-Score are the most relevant performance metrics for event detection algorithms. These normed metrics allow a general performance conclusion considering the number of correctly detected (true positives), incorrectly detected (false positives) and not detected (false negatives) events. The F-Score is a metric that rises to 1 by an increase of true positives and decrease of false positives. It is combining both relevant performance metrics (true positives and false positives) and is the preferred performance metric in the following evaluations.

\begin{align*}
\end{align*}

\subsection{Features}
For our first experiment, we implemented the event detection, using adaptive-training and 87 nearest neighbors for the K-NN classifier. These values seemed promising in pre-executed experiments. The highest performance could be achieved with features that are based on the CUSUM (see Table~\ref{tab:featureResults}). The CUSUM has already been used for event detection with promising results by \citet{Trung2014}. Since the current and $\Delta$CUSUM segments are similar (see Figure~\ref{fig:eventFeatures}), we expected comparable results. A closer look at the segments reveals that the mean event step of the $\Delta$CUSUM segments is broader and more obvious due to the power neutrality of the CUSUM. We assume that this power neutrality leads to a more distinct event model and an improved detection performance. The performance on BLOND-50 supports these assumptions with a similar trend in the results.

Events that have a previous current of near zero are always ON-events, which are easily detectable in the per-appliance measurements (BLOND-50) but not in the case of concurrent running appliances of aggregated measurements (BLUED). The features \emph{ADM}, \emph{SPF}, and \emph{Current} could therefore not be applied to the BLOND-50 dataset due to their strong dependence on the appliance power in combination with the single appliance measurements which would influence the results in an invalid way.

\begin{table}[h]
\centering
\setlength{\tabcolsep}{1.45mm} 
\caption{Feature Results for BLUED and BLOND-50}
\label{tab:featureResults}
\begin{tabular}{lrrrllll}
                           & \multicolumn{3}{c}{BLUED} &  & \multicolumn{3}{c}{BLOND-50} \\ \hline \hline
Feature                    & Prec.   & Rec.   & F-Sc.  &  & Prec.   & Rec.   & F-Sc.  \\ \hline
$\Delta$\emph{CUSUM}   & 0.81   & 0.75  & 0.78  &  & 0,22   & 0,98  & 0,36        \\
\emph{CUSUM}             & 0.80   & 0.75  & 0.78  &  & 0.23   & 0.98  & 0.38        \\
\emph{Current}           & 0.88   & 0.38  & 0.53  &  & -       & -      & -         \\
\emph{ADM}               & 0.88   & 0.38  & 0.53  &  & -       & -      & -         \\
\emph{SPF}               & 0.87   & 0.28  & 0.43  &  & -       & -      & -         \\
\emph{$\Delta$Current} & 0.20   & 0.33  & 0.25  &  & 0.18   & 0.83  & 0.29        \\ \hline
\end{tabular}
\end{table}

\subsection{Normalization}
To prevent undesired feature weighting, a feature normalization needs to be applied, especially in the case of a strong range variance of the feature dimensions. There are two common ways to normalize the feature space. The first is the min-max scaling that ensures that all dimensions lie in a range of [-1 \dots 1] while the second is called standardization that ensures that the standard deviation of all dimensions lies at exactly 1.

\begin{table}[htbp]
\caption{Normalization Results}
  \begin{tabular}{lrrrclll}
  & \multicolumn{3}{c}{BLUED} &  & \multicolumn{3}{c}{BLOND-50} \\ \hline \hline
  Norm        & Prec.         & Rec.     & F-Sc.     & & Prec. & Rec. & F-Sc. \\ \hline
   \emph{None}             & 0.82         & 0.74     & 0.78     & & 0.22 & 0.98 & 0.36 \\
   \emph{MinMax}         & 0.82         & 0.75     & 0.78     & & 0.23 & 0.97 & 0.37 \\
   \emph{Variance}     & 0.82         & 0.72     & 0.77     & & 0.24 & 0.96 & 0.38 \\ \hline
  \end{tabular}
  \label{tab:normResults}
\end{table}

The min-max normalization performs best in our experiments on BLUED but also shows that the normalization itself does not influence the performance significantly (see Table~\ref{tab:normResults}). For BLOND-50, the best result could be achieved with a variance normalization. However, also here, the performance results remain quite stable. This means that the different value ranges of the feature space dimensions do not add any significant weighting. This is most likely caused by a similar order of magnitude in the value range across the individual feature space dimensions. The fact that the features are based on time series segments, and therefore share the same value range, affirms the low variations in the performance results.


\subsection{Training Method}
The two previously introduced training methods (classical and adaptive training) are being evaluated. The best result for the multivariate event detection (without adaptive training) allows detection of 1\,170 out of 1\,577 appliance events from BLUED with 490 false positives and a corresponding F-Score of 0.72. This result was obtained with 30 periods of step-size and the K-NN classifier with K=301.

\begin{table}[htbp]
\centering
\caption{Adaptive Training Improvement on BLOND-50}
\label{tab:AdaptImprovement}
\begin{tabular}{lrrrclll}
& \multicolumn{3}{c}{K-NN} &  & \multicolumn{3}{c}{SVM} \\ \hline \hline
Training & Prec. & Rec.  & F-Sc. && Prec. & Rec.  & F-Sc.      \\ \hline
\emph{classical}   & 0.13 & 0.99 & 0.24 && 0.12 & 0.99 & 0.21  \\
\emph{adaptive}    & 0.22 & 0.98 & 0.36 && 0.28 & 0.94 & 0.43  \\
\emph{adaptive 3x} & 0.45 & 0.87 & 0.59 && 0.55 & 0.85 & 0.67  \\
\emph{adaptive 5x} & 0.53 & 0.85 & 0.65 && 0.56 & 0.77 & 0.65  \\ \hline
\end{tabular}
\end{table}

\begin{figure}[htbp]
\centering
\includegraphics[width=0.65\linewidth]{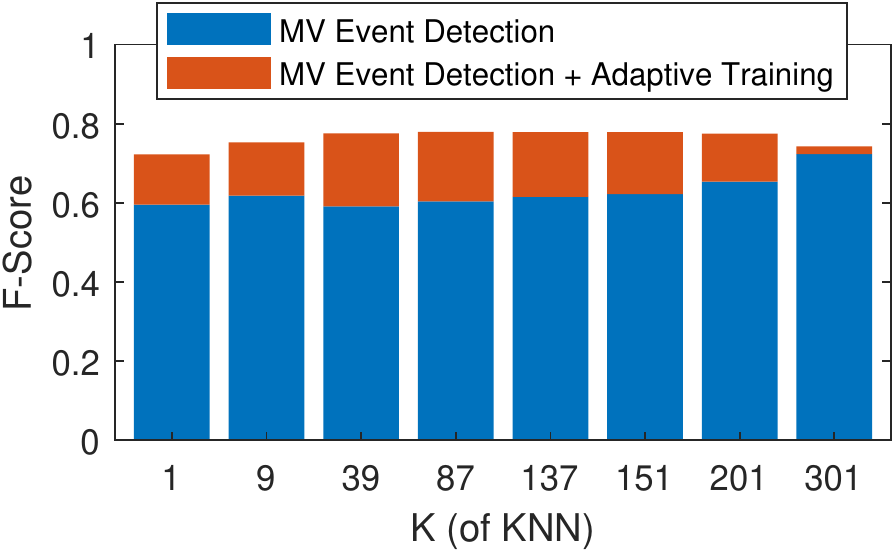}
\caption{The individual performance improvements by using the adaptive training. The bars show the achieved event detection F-Score for different K of the K-NN classifier on the BLUED dataset.}
\label{fig:AdaptTrainImprovement}
\Description{Bar chart of adaptive training improvements.}
\end{figure}

All experiments for the adaptive training show a significant, absolute improvement of the event detection performance of +0.14 in average for the F-Score regarding the BLUED dataset (see Figure~\ref{fig:AdaptTrainImprovement}). The individual improvements vary slightly. The primary performance enhancement of the adaptive training is to reduce the number of false positives due to improvements in the non-event class. The best result for BLUED was obtained with 1\,175 true positives and an F-Score of 0.78 by using K=137 for the K-NN classifier, a min-max normalization, and one adaptive training round. The number of false positives was reduced to 260. A significant rise of true positives was not expected and did not occur in most experiments with adaptive training.

\begin{figure}[tbp]
\centering
{\includegraphics[width=0.95\linewidth]{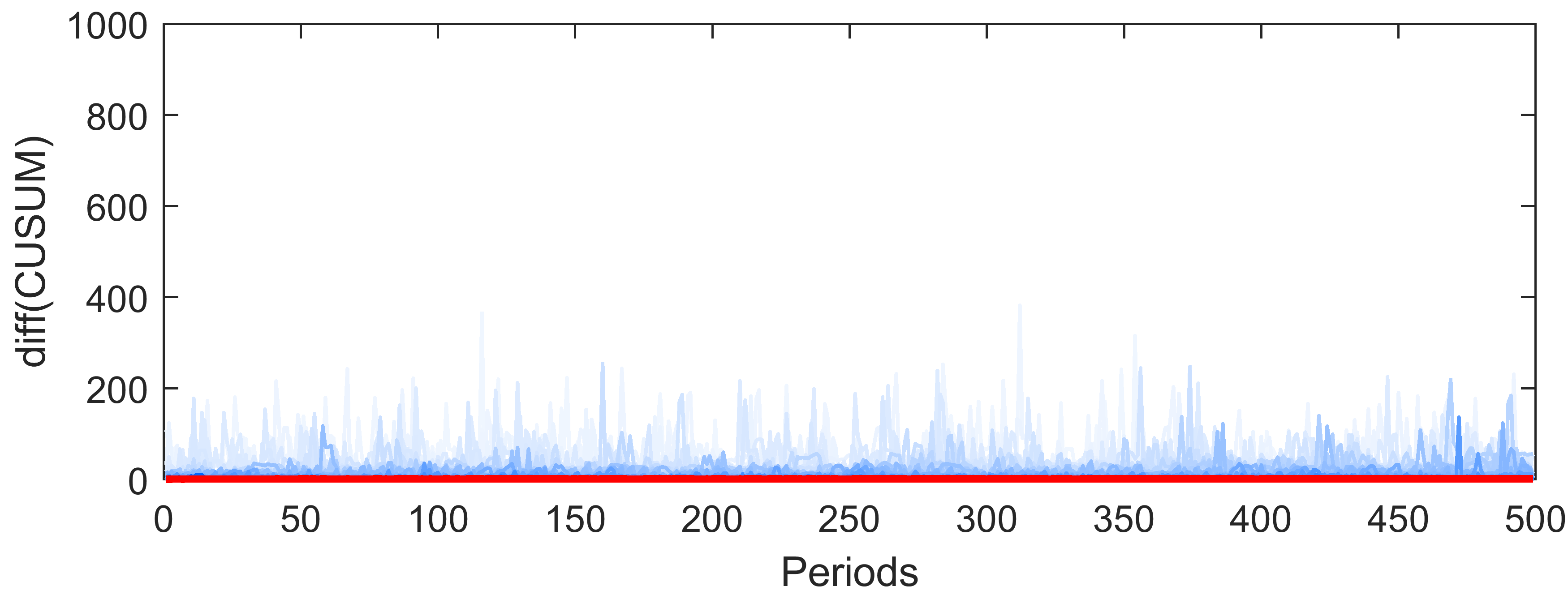}}
{\includegraphics[width=0.95\linewidth]{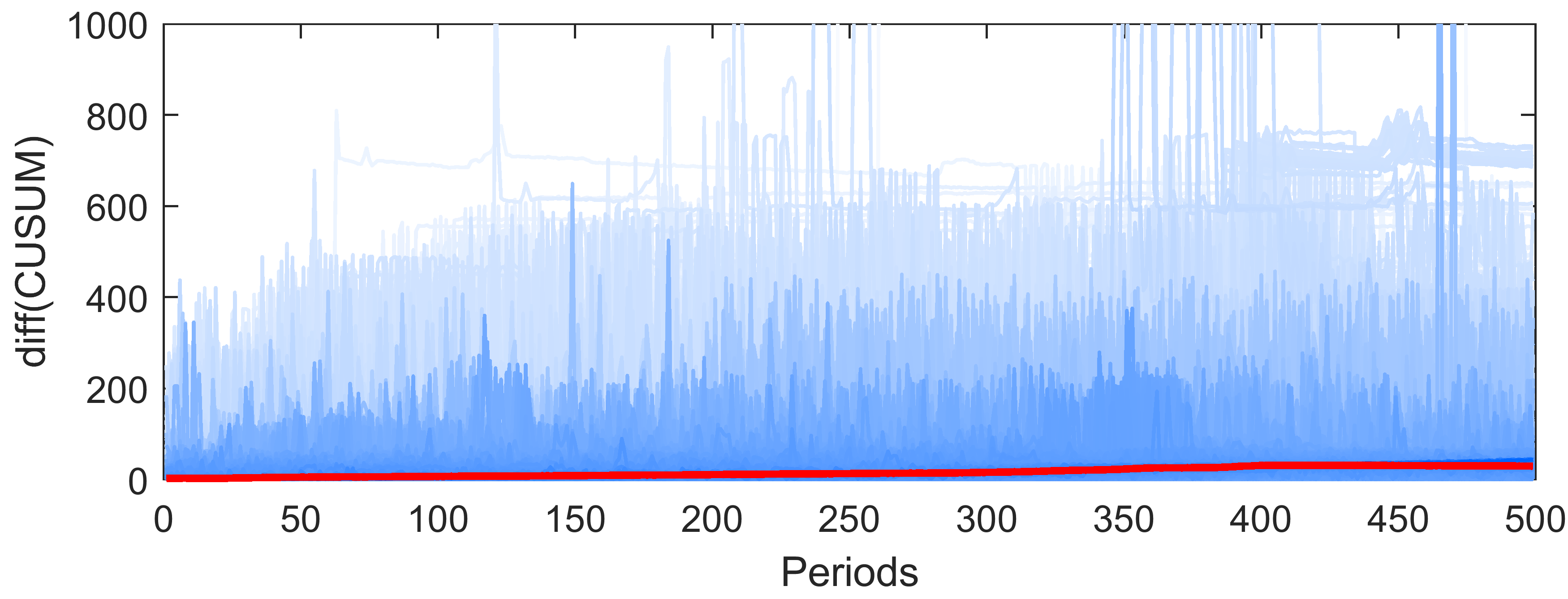}}
\caption{The first plot shows the non-event class represented only by the implicitly-known non-event segments. The second plot shows the non-event class including the false positives from the adaptive training. The increased diversity due to the false positives is clearly visible. The images are retrieved from non-events of the first 2 weeks in 2016-11 of the BLOND-50 dataset without (first plot) and with (second plot) one adaptive training run.}
\label{fig:non-eventComparison}
\Description{Plot of all non-events vs. plot of all non-events including false positives from adaptive training.}
\end{figure}

The main improvement was observed by applying three rounds of the adaptive training to the event detection on the BLOND-50 dataset. Since the event detection on this dataset produces many false positives, due to a high number of SMPS-driven appliances, the adaptive training reduced the number of false positives from 19\,463 to 2\,297 which is an improvement of more than eight times. An expected side effect of this enormous improvement is a considerable, but still low, decrease in true positives and recall (see Table~\ref{tab:AdaptImprovement}).

\begin{table}[htbp]
\centering
\small
\setlength{\tabcolsep}{1.15mm} 
\caption{Overall best results on BLUED and BLOND-50}
\label{tab:overallBestResults}
\begin{tabular}{lcccccc}
         & Feature       & Norm     & Train   & Class & Param       & F-Sc. \\ \hline \hline
BLUED    & $\Delta$CUSUM & MinMax   & adap 1x & KNN   & K=137       & 0.78    \\
BLOND-50 & CUSUM         & Variance & adap 3x & SVM   & C/G 128/512 & 0.67    \\ \hline
\end{tabular}
\end{table}


Using the adaptive training to augment the training set with false positive samples, we were able to reduce the final number of false positives during testing. We conclude that the classifier learns the not explicitly definable heterogeneous model of a non-event by adding the false positives of the training run (see Figure~\ref{fig:non-eventComparison}).

\begin{figure}[tbp]
\vspace{-3mm}
\minipage{\linewidth}
  \centering
  \includegraphics[width=\linewidth]{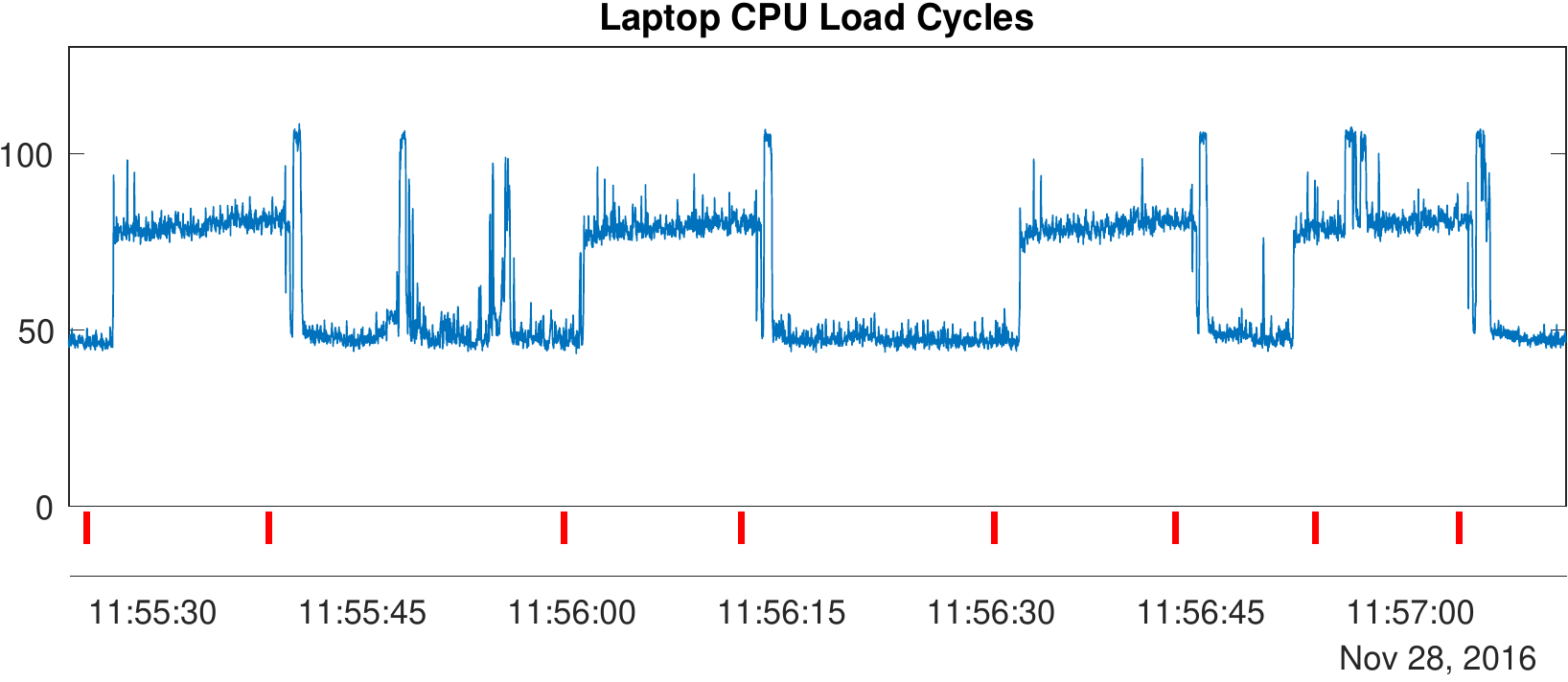}
  \label{fig:fpLaptop}
\endminipage\hfill
\vspace{-1mm}
\minipage{\linewidth}
  \centering
  \includegraphics[width=\linewidth]{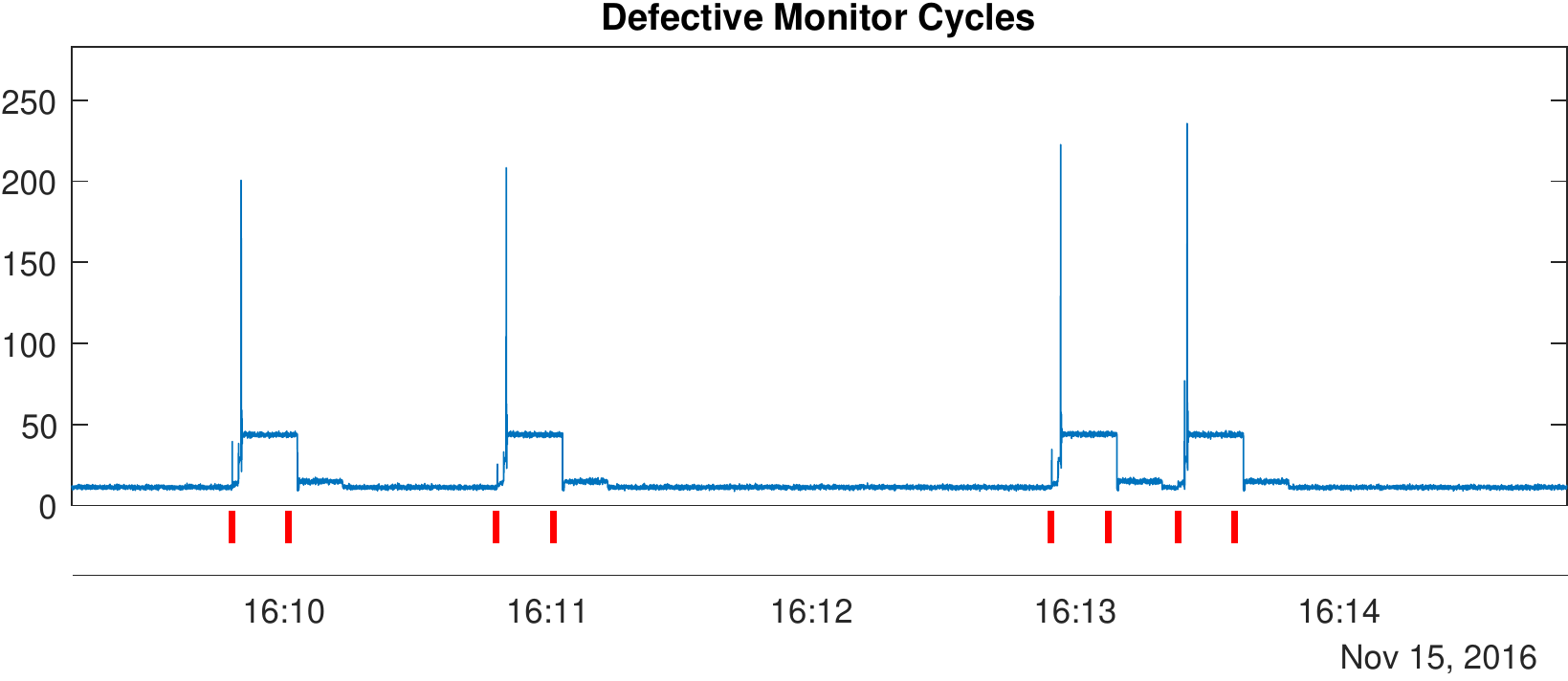}
  \label{fig:fpMonitor}
\endminipage\hfill
\vspace{-1mm}
\minipage{\linewidth}%
  \hspace{0.3mm}
  \includegraphics[width=\linewidth]{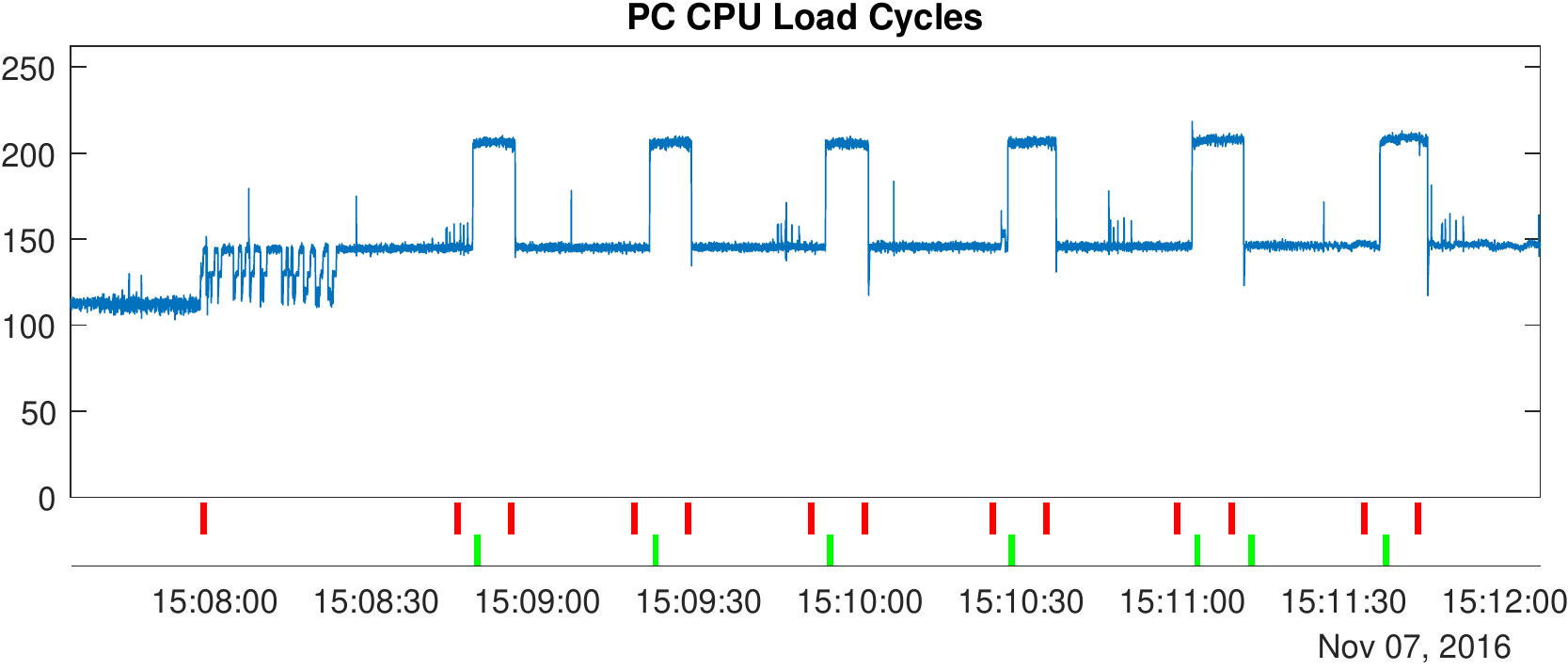}
  \label{fig:fpPC}
\endminipage
\vspace{-2mm}
\caption{The three plots show the most prominent reasons for false positive events in BLOND-50: a laptop that produces event-like patterns (first plot), a faulty monitor that immediately goes OFF after switching ON (second plot), and a desktop computer that produces event-like patterns due to CPU load changes. The colored event marker show the false positives that stem from the classical (red) and adaptive (5x) training method (green).}
\label{fig:fpAppliances}
\Description{Three plots that show the most prominent reasons for false positives.}
\end{figure}



\subsection{Classification}

\subsubsection*{K-NN}
Since the event detection performance varies unexpectedly strong, depending on the number of neighbors for the K-NN classifier, we decided to evaluate the performance of eight different K for the classifier. The best general K in our experiments was 301 with classical training, while it was 137 when applying the adaptive training (see Figure~\ref{fig:AdaptTrainImprovement}). For BLOND-50 the best result with K-NN was achieved by using five rounds of adaptive training (see Figure~\ref{fig:fpAppliances}).



\subsubsection*{SVM}
The best result we could achieve by using the SVM classifier on the BLUED dataset was with an F-Score of 0.72 considerably lower than with 0.78 for the K-NN classifier. The reason is an almost twice the number of false positives - even after adaptive training. The number of true positives with 1\,112 lies only slightly below the best result for K-NN. For BLOND-50 the best result by using the SVM lies in a range of 0.67 by using three adaptive training rounds. The optimal SVM hyper-parameter have been retrieved with a grid search algorithm that is provided in the LIBSVM package of \citet{Chang2011} and could be found at C=128 and Gamma=512 for BLUED and C=1 and Gamma=0.0078 for BLOND-50.


%


\section{Conclusions}
We propose a multivariate event detection that learns from a user designed event model. The event model stems from event and non-event segments of the training area and allows a user relevant event detection. The challenge to distinguish between relevant and irrelevant events is tackled by multiple runs of an introduced adaptive training process. For events of the BLUED dataset, an F-Score of 0.78 could be achieved, which lies in a range of the state-of-the-art. It allows a reduction of more than eight times of false positives for BLOND-50. We could achieve an F-Score of 0.67, which means that a found event is more likely relevant than irrelevant for the user.

The multivariate event detection in combination with the introduced way of adaptive training is an appropriate algorithm for the increasing number of SMPS-driven appliances in residential and office environments.



\begin{acks}
 We would like to thank Hardani Maulana for the thorough event annotation on the BLOND-50 dataset. This research was partially funded by the Alexander von Humboldt Foundation established by the government of the Federal Republic of Germany and was supported by the Federal Ministry for Economic Affairs and Energy on the basis of a decision by the German Bundestag.
\end{acks}


\balance
\bibliographystyle{ACM-Reference-Format}
\bibliography{bibliography}

\end{document}